\documentclass[12pt,preprint]{aastex}
\usepackage{epsfig}

\shorttitle{CXOGC~J174536.1-285638: X-ray and Infrared Variability}
\shortauthors{Mikles et al.}

\begin{document}

\title{Discovery and Interpretation of an X-ray Period in the Galactic Center Source CXOGC~J174536.1-285638}
\author{Valerie J. Mikles}
\affil{Department of Astronomy, University of Florida, Gainesville, FL 32611}
\email{mikles@astro.ufl.edu}

\author{Stephen S. Eikenberry}
\affil{Department of Astronomy, University of Florida, Gainesville, FL 32611}
\email{eikenberry@astro.ufl.edu}

\author{Reba M. Bandyopadhyay}
\affil{Department of Astronomy, University of Florida, Gainesville, FL 32611}
\email{reba@astro.ufl.edu}

\and 

\author{Michael P. Muno}
\affil{Space Radiation Laboratory, Caltech, Pasadena, CA 91125}
\email{mmuno@srl.caltech.edu}

\begin{abstract}

We present X-ray and infrared observations of the X-ray source CXOGC~J174536.1-285638. Previous observations suggest that this source may be an accreting binary with a high-mass donor (HMXB) or a colliding wind binary (CWB). Based on the {\it Chandra} and {\it XMM-Newton} light curve, we have found an apparent $189 \pm 6$ day periodicity with better than 99.997\% confidence. We discuss several possible causes of this periodicity, including both orbital and superorbital interpretations. We explore in detail the possibility that the X-ray modulation is related to an orbital period and discuss the implications for two scenarios; one in which the variability is caused by obscuration of the X-ray source by a stellar wind, and the other in which it is caused by an eclipse of the X-ray source. We find that in the first case, CXOGC~J174536.1-285638 is consistent with both CWB and HMXB interpretations, but in the second, CXOGC~J174536.1-285638 is more likely a HMXB.
\end{abstract}


\section{Introduction}

{\it Chandra} observations of the Galactic Center (GC) have revealed a large new population of low luminosity X-ray sources with $L_X[D/8kpc]^2 \sim 10^{31} - 10^{35} erg~s^{-1}$ \citep{mun03}. In addition, the Swift and INTEGRAL missions have recently revealed a new population of highly-absorbed X-ray sources, believed to be high-mass X-ray binaries \citep[HMXBs; e.g.][]{beck05,bod07,neg07}. In 2005, we identified an infrared (IR) star as the first spectroscopically confirmed IR counterpart to the low luminosity {\it Chandra} source CXOGC~J174536.1-285638 \citep[hereafter Paper I]{mikles06}. Based on the X-ray and IR spectra and the X-ray to IR luminosity ratio, we showed that the source is most likely a massive star in a binary system. The source shows strong HeI (2.114-$\mu$m), Brackett-$\delta$ (1.945-$\mu$m), and Brackett-$\gamma$ (2.166-$\mu$m) emission lines, typical of both accretion-powered binaries and CWBs. Additionally, we observe Brackett series, HeI, HeII, CIII, and NIII line emission. P~Cygni profiles are visible in several HeII lines, suggesting wind activity around a massive star. The X-ray spectrum of this source is particularly intriguing, having prominent Fe-XXV emission centered at 6.7~keV with an equivalent width of 2.2~keV. This is one of the highest equivalent width Fe-XXV lines ever seen (Paper I). 

Our initial IR observations were aimed toward the discovery of a short ($<1$~d) period in the CXOGC~J174536.1-2856 binary. We use IR spectroscopy to search for variations in CXOGC~J174536.1-285638's IR emission features. We analyze {\it Chandra} and {\it XMM-Newton} archival data to search for X-ray variability over short and long baselines. From the combined X-ray light curve, we find a period of $189 \pm 6$ days. We discuss CXOGC~J174536.1-2856's variability in the IR and X-ray, and examine the implications of a 189~d period for the nature of the source. In \S 2, we summarize our IR and X-ray observations and analysis, detailing both our IR radial velocity study and X-ray period analysis. In \S 3, we discuss the IR and X-ray variability in CXOGC~J174536.1-2856, specifically exploring an orbital period interpretation of the identified X-ray period. 

\section{Observations and Analysis}

\subsection{Infrared Counterpart to CXOGC~J 174536.1-285638}

CXOGC~J174536.1-285638 was discovered as part of a {\it Chandra} survey of the GC region. The survey, conducted by \citet{mun03}, identified 2357 serendipitous X-ray sources with $L_X[D/8kpc]^2 \sim 10^{31} - 10^{35} erg s^{-1}$ within $\sim 10$-arcmin of Sgr A*. The coordinates of the source are 266.40060, -28.94407 with positional uncertainty of 0.4-arcsec \citep{mun04}.

\begin{figure*}
\epsscale{0.8}
\plottwo{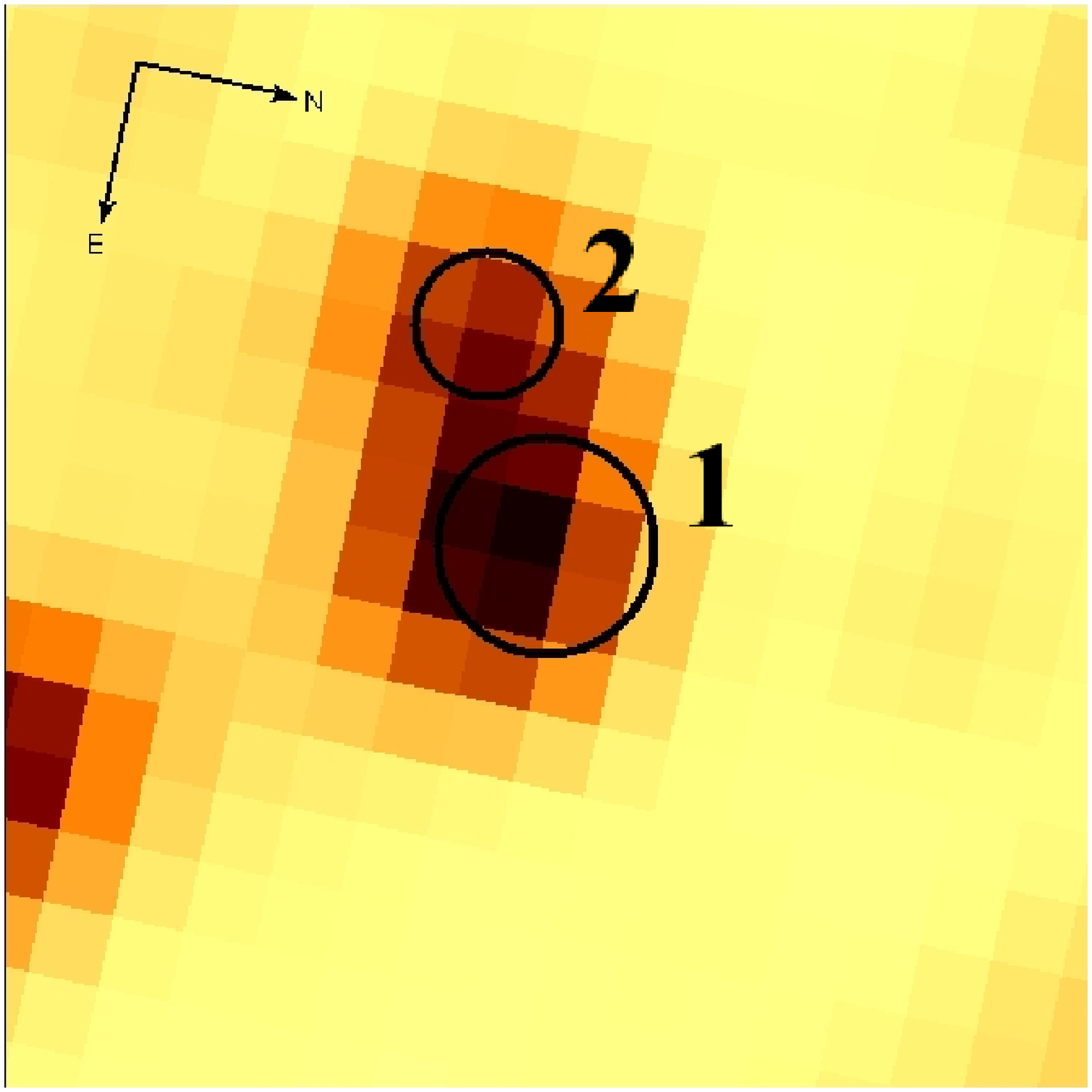}{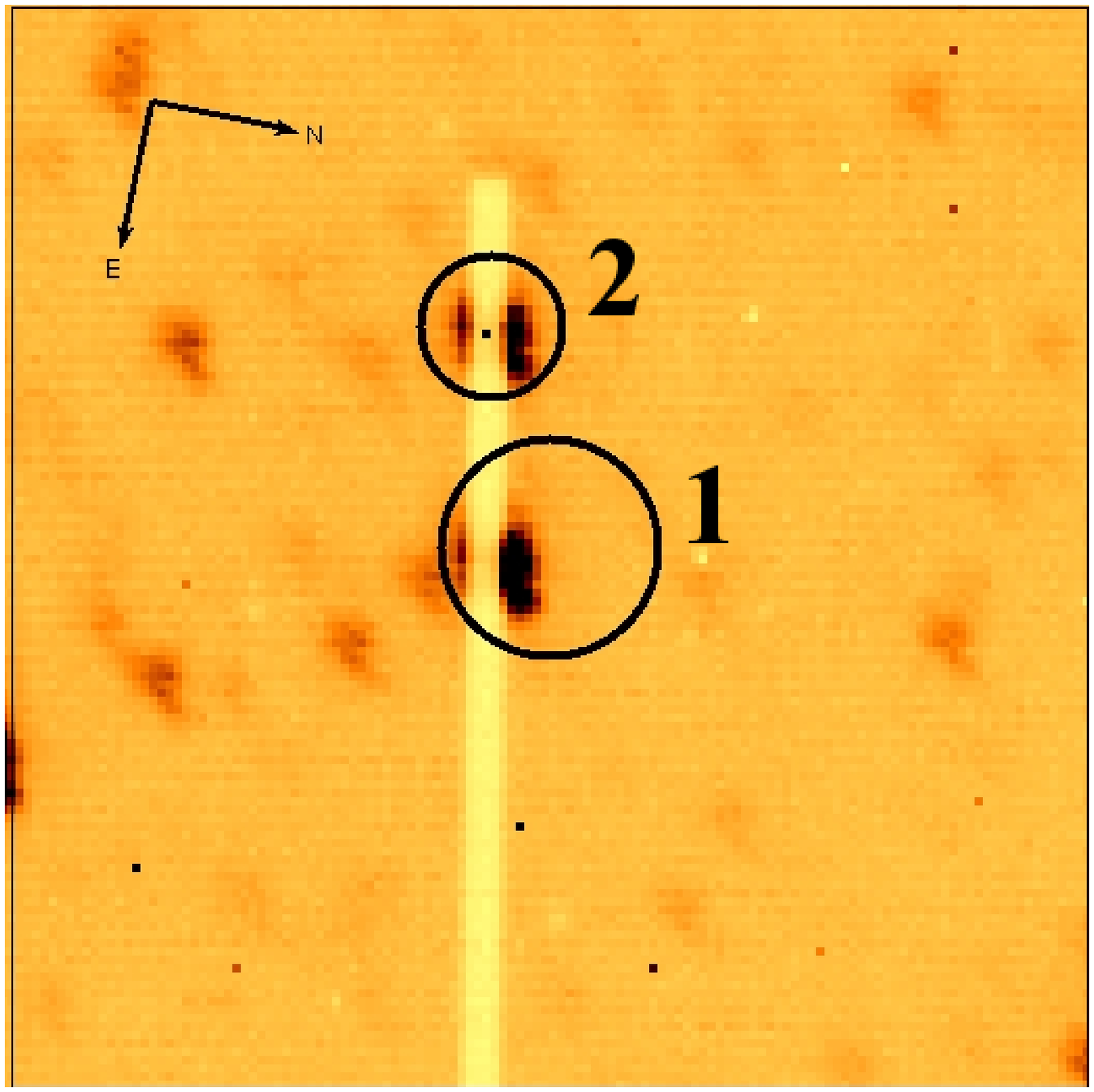}
\caption{Left: A 15'' x 15'' image of the 2MASS region near the {\it Chandra} X-ray coordinate center.  A 1.5 arcsec circle is drawn around the {\it Chandra} source coordinates. A second circle is drawn around the blended source. Right: A 15'' x 15'' IRTF slit image of the same region. The stars blended in the 2MASS region are clearly resolved on the slit.}
\label{fig1}
\end{figure*}

\begin{figure*}
\plotone{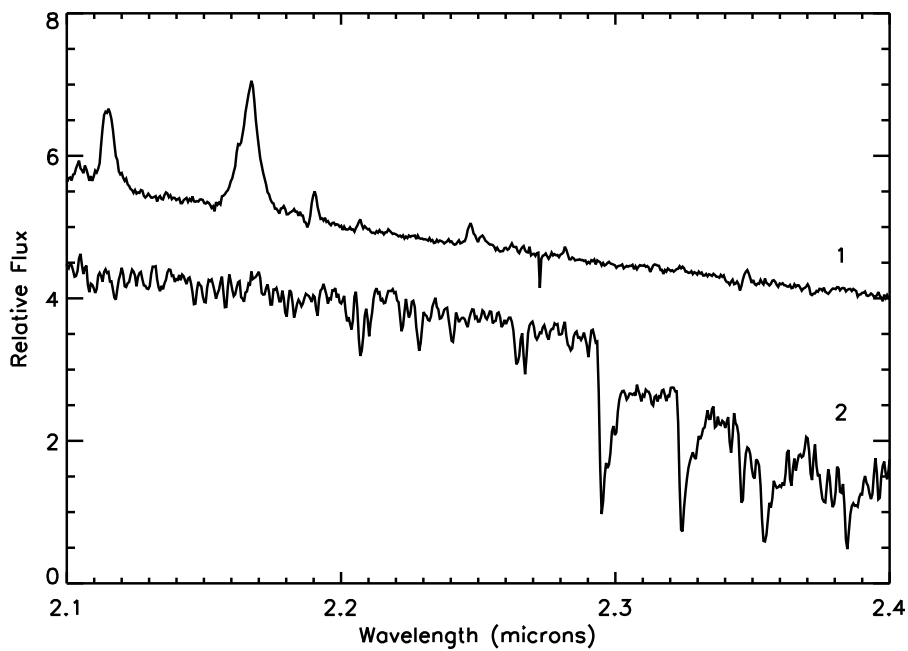}
\caption{K-band spectra of CXOGC~J174536.1-285638 and the neighbor star. The two objects are blended in 2MASS, but clearly resolved by IRTF. Source 1 is the likely X-ray counterpart. Source 2 is a type K or cooler evolved source, lacking emission lines which would be indicative of energetic processes.}
\label{fig2}
\end{figure*}

We searched for potential IR counterparts using the 2MASS catalog and identified the 2MASS source 17453612-2856386 as the likely counterpart. The blended 2MASS source is clearly resolved into two stars in our IRTF observations. In Figure  \ref{fig1}, we show a 15'' x 15''  2MASS image and IRTF SpeX slit image with a 1.5'' circle at the {\it Chandra} coordinate center. The two stars, blended in 2MASS, are well separated in the IRTF finder image. The IR astrometric solution is derived from 2MASS which has a stated astrometric accuracy of 15mas. Due to the proximity of the two potential IR counterparts, we were able to obtain simultaneous spectra of both stars; we plot both spectra in Figure \ref{fig2}. In Paper I, we identify ``star 1,'' the emission line source, as ``Edd-1,'' the counterpart to the {\it Chandra} object. The second source is an evolved star of type K or cooler, with no evidence for emission lines which are signatures of high energy processes, such as accretion or wind collison. It thus seems unlikely that this second source is the IR counterpart to the X-ray source.  

\subsection{Search for short-period IR variability}

On 2006 Aug 02-04 UT, we obtained J, H, and K band (1.1-2.4 $\mu$m) spectra of the IR counterpart to CXOGC~J174536.1-285638 using SpeX on IRTF \citep{rayner03}, in hopes of finding short period ($<1$~d) variability in the source. Dithering along the 0.5 arcsec slit, we obtained 184 exposures of 120~s each over the course of three half-nights, giving us a time baseline of 3-4 hours per night. The procedure for our IR analysis of the SpeX data is described in Paper I. We extract spectra using the standard SpexTool procedure for AB nodded data, resulting in a series of sky-subtracted, wavelength-calibrated spectra \citep{vacca03,cushing04}. We interpolate over the intrinsic Brackett absorption features in the G0V-star spectrum, then divide the target spectrum by the G0V-star in order to remove atmospheric absorption bands. We multiply the resultant spectrum by a 5900~K blackbody spectrum, corresponding to the temperature of the G0V-star.  
Using our previous observations taken on 2005 July 1 UT (Paper I), we adopt a reddening value of $A_V=29$ mag and apply this correction to all data. 

\begin{figure*}
\plotone{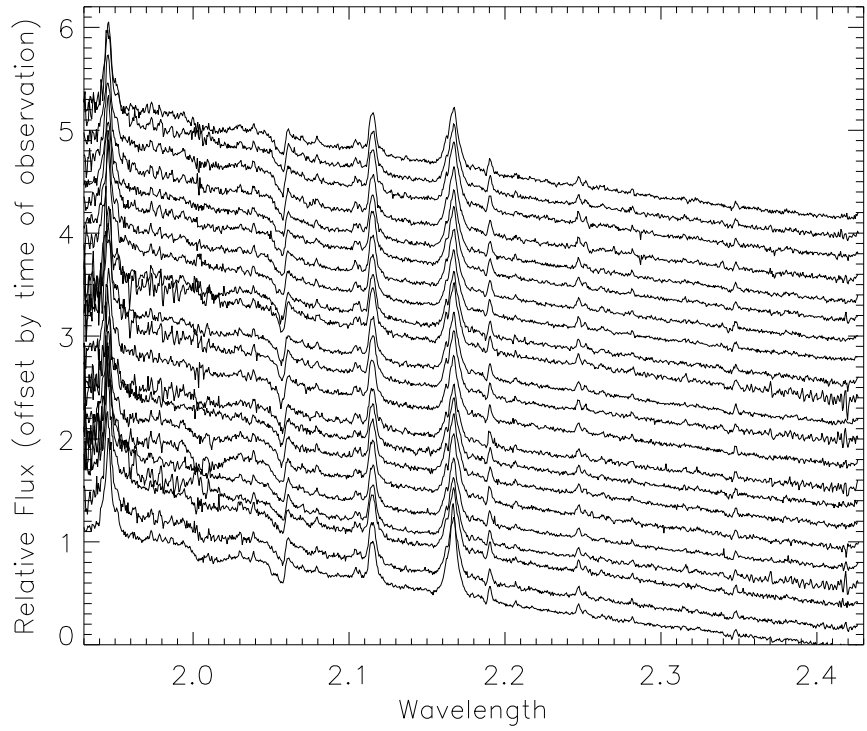}
\caption{The K-band spectra of CXOGC~J174536.1-285638. We show the original 2005 spectrum at the bottom and the twenty-minute combinations of the 2006 spectra over the three nights. These are offset by time of observation, such that the earliest spectra are lower and later are higher. The relative times of these spectra are listed in Table \ref{tbl-1}.}
\label{fig3}
\end{figure*}

\begin{figure*}
\epsscale{0.8}
\plotone{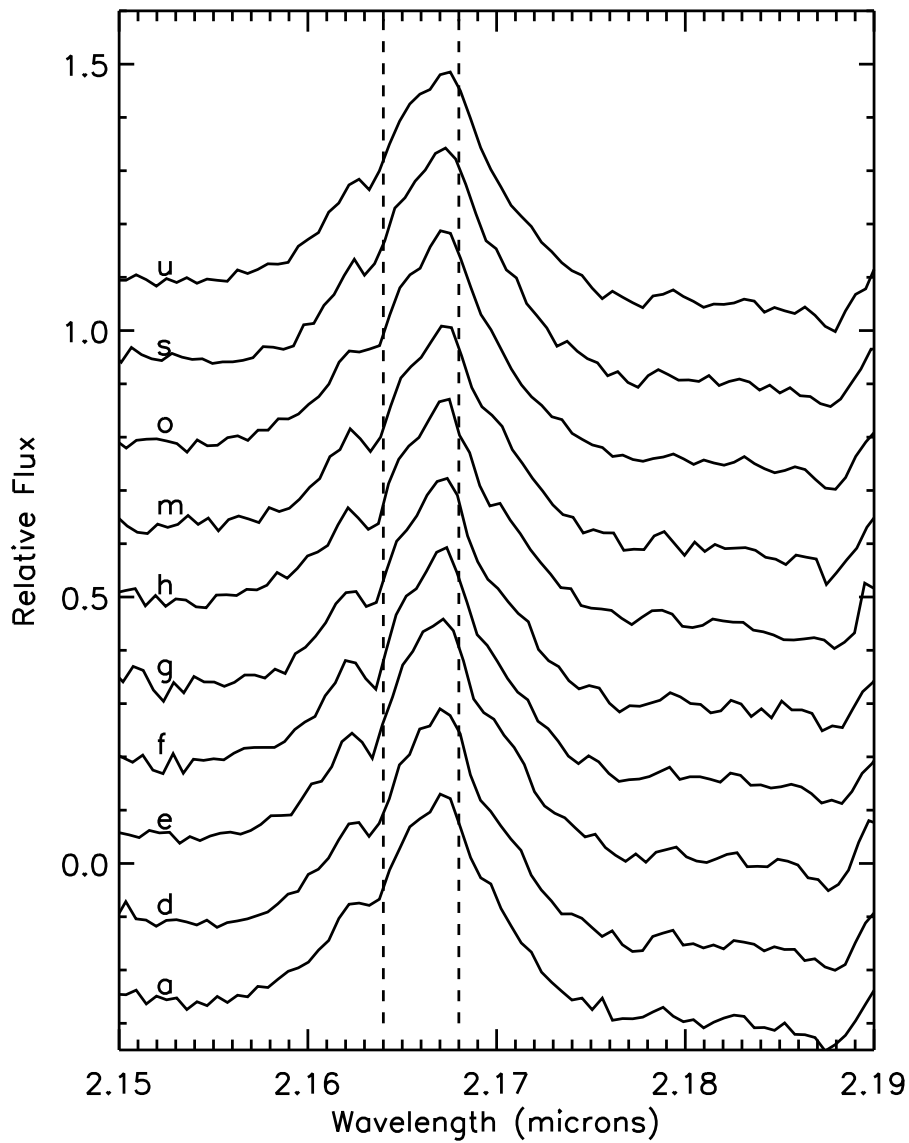}
\caption{The Br-$\gamma$ region of select CXOGC~J174536.1-285638 spectra taken from 2006 Aug 02-04. The region shows apparent non-periodic variation, mostly around the $2.164 \mu$m Helium contribution. These variations are only occasionally greater than 5-times the RMS spectral difference. Higher resolution spectroscopy is needed to show whether this is intrinsic to CXOGC~J174536.1-285638 or an artifact of the data reduction. The relative times of these spectra are listed in Table \ref{tbl-1}.}
\label{fig4}
\end{figure*}

We used spectra from each night to test variability on multiple time scales. Figure \ref{fig3} shows the series of twenty-one K-band spectra taken over the course of our observations, with integration times between 8 and 20 minutes per spectrum. We list the specific time stamps and exposure times of these spectra in Table \ref{tbl-1}. To search for radial velocity variations in the emission lines, we track the line centers with two methods: first by taking a statistical mean of the wavelength around the line center, weighted by flux, and second by fitting a Gaussian to the line. We find no radial velocity variations, nor do we find significant flux variations in the lines. We checked for IR variability on 1 year, 3 day, 3 hour, 1 hour, and 30 minute baselines and found no evidence of periodic variability or flares in this sample.  The only apparent variation is in the structure of the Br-$\gamma$ line complex (see Figure \ref{fig4}), but this does not often vary more than $\sim 5$ times the RMS spectral difference in the vicinity of the $\lambda 2.164$ Helium component. Further we note that this region is affected by our data reduction process (i.e., the removal of the intrinsic Brackett absorption in the G0V).

\subsection{X-ray Variability}

\begin{figure*}
\epsscale{0.8}
\plotone{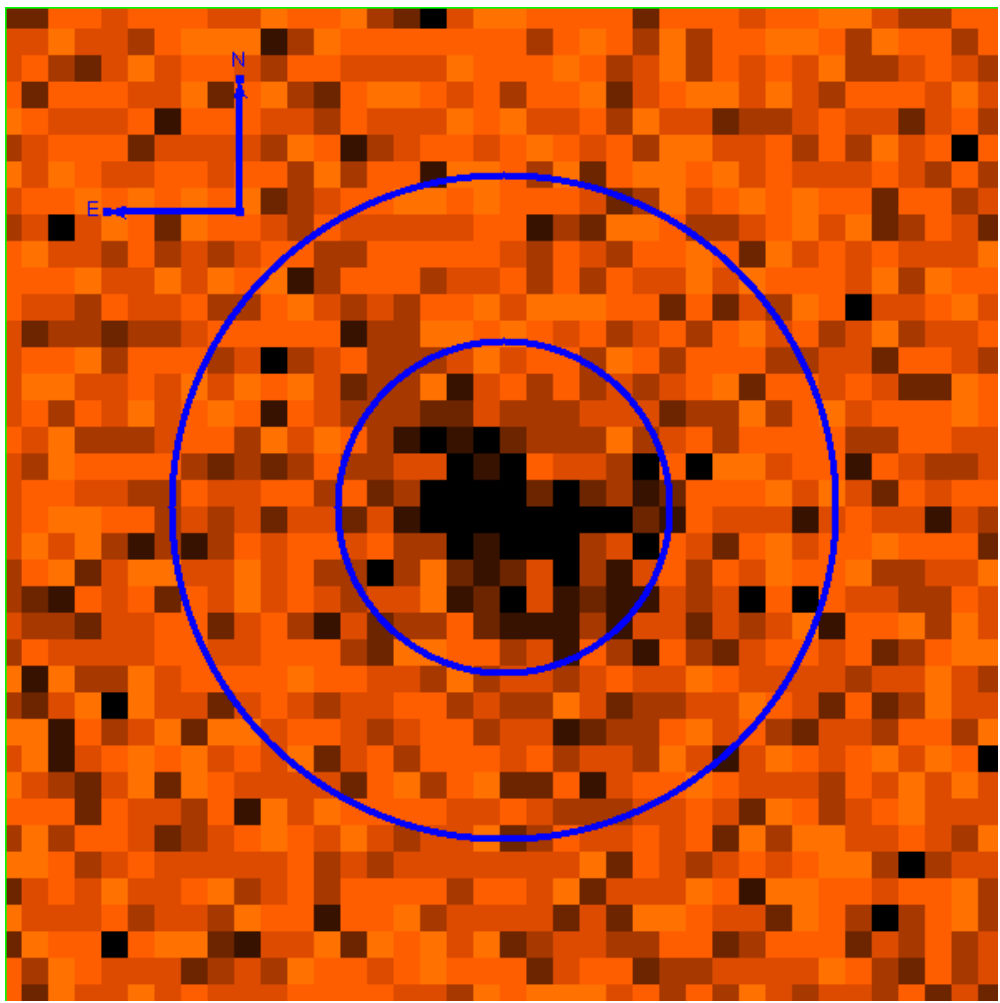}
\caption{A 60'' x 60'' {\it XMM} image centered around the {\it Chandra} source coordinates (denoted by the inner circle). The concentric circles denote the region of source counts and background counts used in analysis of the {\it XMM} data.}
\label{fig5}
\end{figure*}

\begin{figure*}
\includegraphics[angle=90,width=\textwidth]{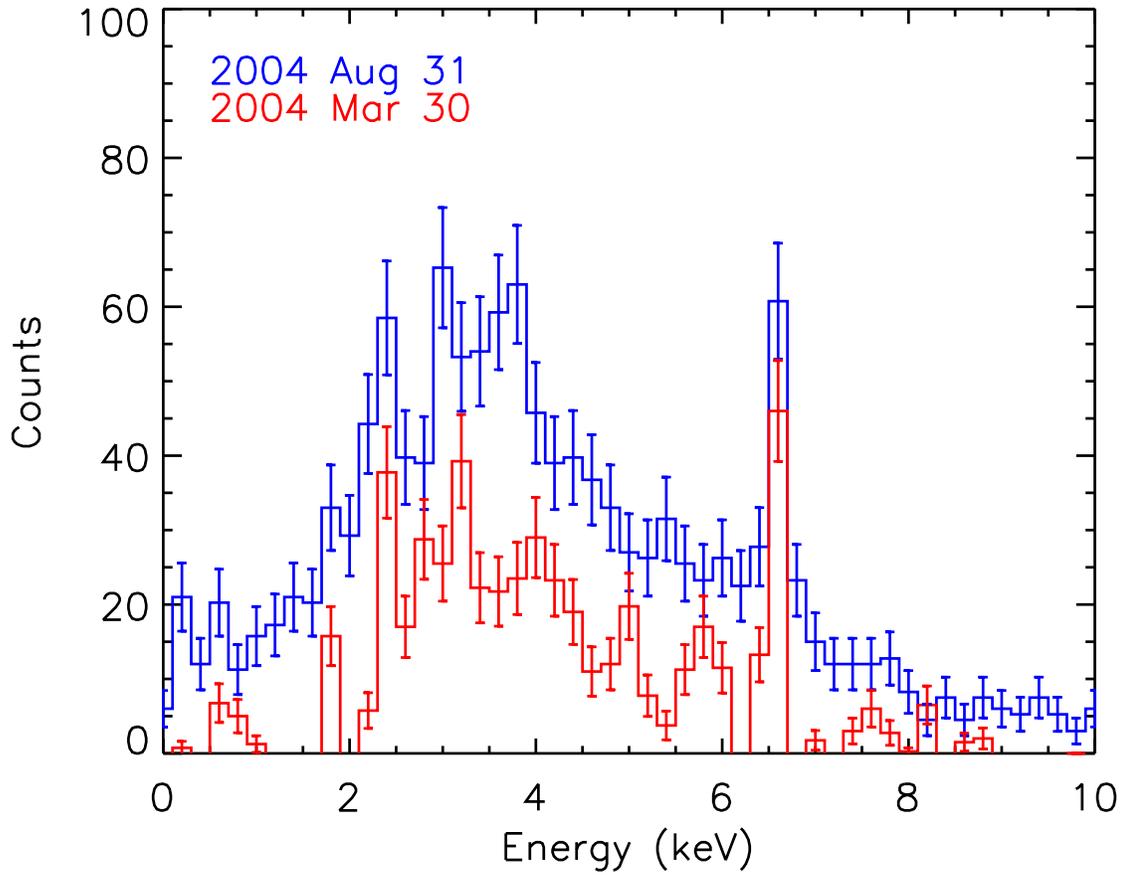}
\caption{Two representative {\it XMM} spectra separated by 0.7 in phase. While the strength of the FeXXV line is consistent between the two observations, the continuum level drops significantly. If such variation were caused entirely by column absorption due to a stellar wind, then $N_H$ would increase by $2.5 \times 10^{23} cm^{-2}$.}
\label{fig6}
\end{figure*}

{\it Chandra} observations of CXOGC~J174536.1-285638 revealed long baseline intensity variations by a factor of $\sim$3 in X-ray in the 2-8~keV range. The variation, which was observed initially with {\it Chandra} in 2002 as a drop in flux, repeated in 2006 with similar morphology, prompting us to search archival X-ray data for additional information about this source's long term (month to year) variability. We list the {\it Chandra} data used in our analysis in Table \ref{tbl-2}. We supplement the {\it Chandra} observations with {\it XMM-Newton} archival data, listed in Table \ref{tbl-3}. CXOGC~J174536.1-285638 is easily identified as an isolated source $\sim$10~pc from the GC and is not confused with any other source detection in either the {\it Chandra} or {\it XMM} images. The positions of the {\it XMM} and {\it Chandra} sources are consistent within the respective astrometric accuracy of the two instruments \citep[0.4'' and 1'' respectively;][]{mun04b, kirsch05}. We identify the {\it XMM} counterpart to the {\it Chandra} source and show that in Figure \ref{fig5}. In addition to the astrometric accuracy, strong Fe-emission is detected in both the {\it Chandra} and {\it XMM} data (see Fig. \ref{fig6}), confirming that the {\it XMM} source is the same as the {\it Chandra} source. The Fe-XXV emission in this source is unusually strong and it would be extremely unlikely to detect emission in both the {\it Chandra} (see Paper 1) and {\it XMM} (this paper) spectra were they not the same source.  

Due to the relative faintness of CXOGC~J174536.1-285638 in the X-ray (usually $<$20 cts/hr), many of the {\it XMM} observations suffer from low signal-to-noise. While the 2001-2002 data consist of fairly short observations (exposure time $< 7$~h), in 2004 there are four observations of 40 consecutive hours each. Following standard {\it XMM} data reduction techniques, we generate an astrometrically calibrated event list. From this, we located CXOGC~J174536.1-285638 and extracted light curves and spectra from a circle with a 200-pixel radius. The background was calculated from a ring extending 300-500 pixels from the source center. We show the extraction region around the source in Figure \ref{fig5}. We set the spectral bin size at 200~eV and plot two representative spectra in Figure \ref{fig6}. The flux varies by a factor of three between these observations. Because of the extremely low count rate, we cannot meaningfully constrain the fainter spectrum with XSpec models. 

We extract light curves at five hour intervals over the full 2-8~keV band, as well as from the ``soft'' 2-4~keV band and the ``hard'' 4-8~keV band separately for the 2004 {\it XMM} observations. The time resolution is chosen to ensure sufficient counts in each bin to test for variability. The X-ray flux varies aperiodically by less than a factor of 2 over the course of each individual observation and the hardness ratio is consistent with zero. Aperiodic variability is not uncommon in stellar X-ray sources on these timescales. We observe no periodic variability on timescales less than 40 hours. 

\begin{figure*}
\epsscale{0.8}
\plotone{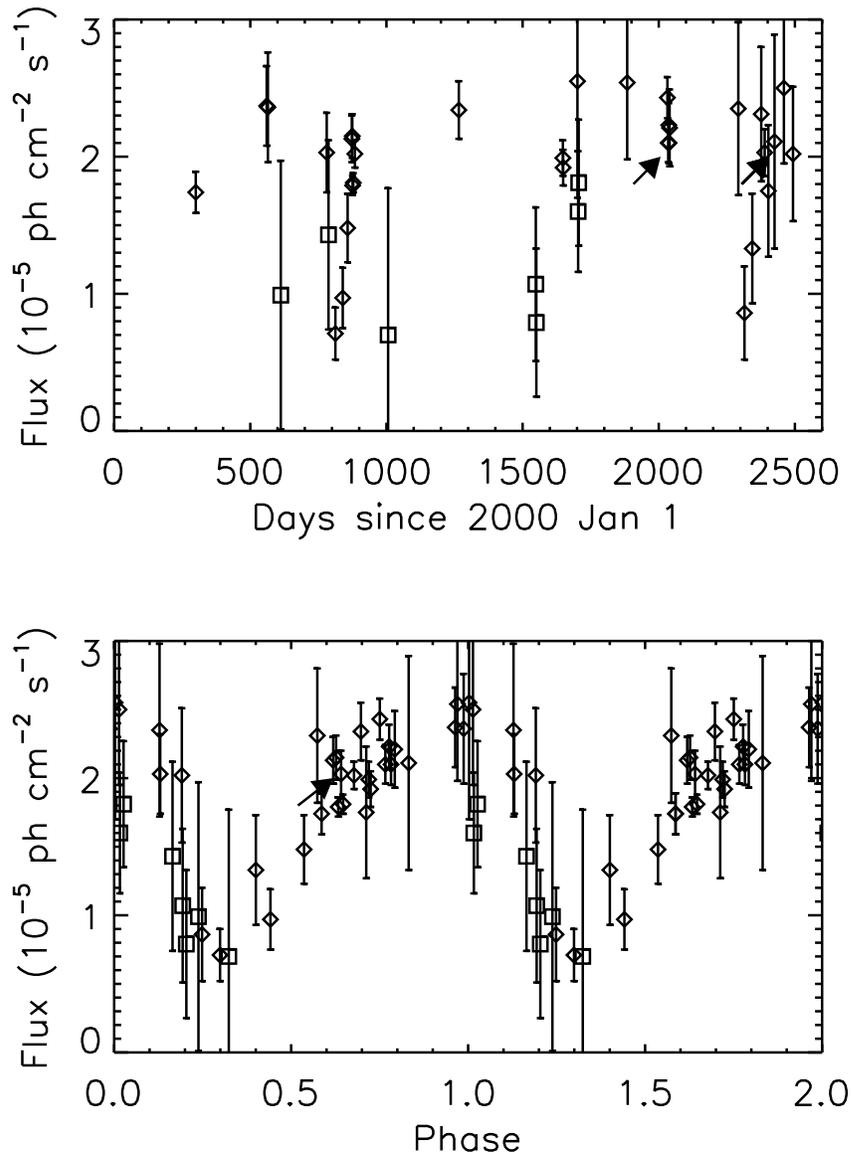}
\caption{The X-ray light curve (top) and folded light curve (bottom) of CXOGC~J174536.1-285638. The light curve is folded on a 189 d period. The squares are {\it XMM} data; the diamonds are {\it Chandra} data. The arrow indicates the data of the IR spectra.}
\label{fig7}
\end{figure*}

\begin{figure*}
\includegraphics[width=\textwidth]{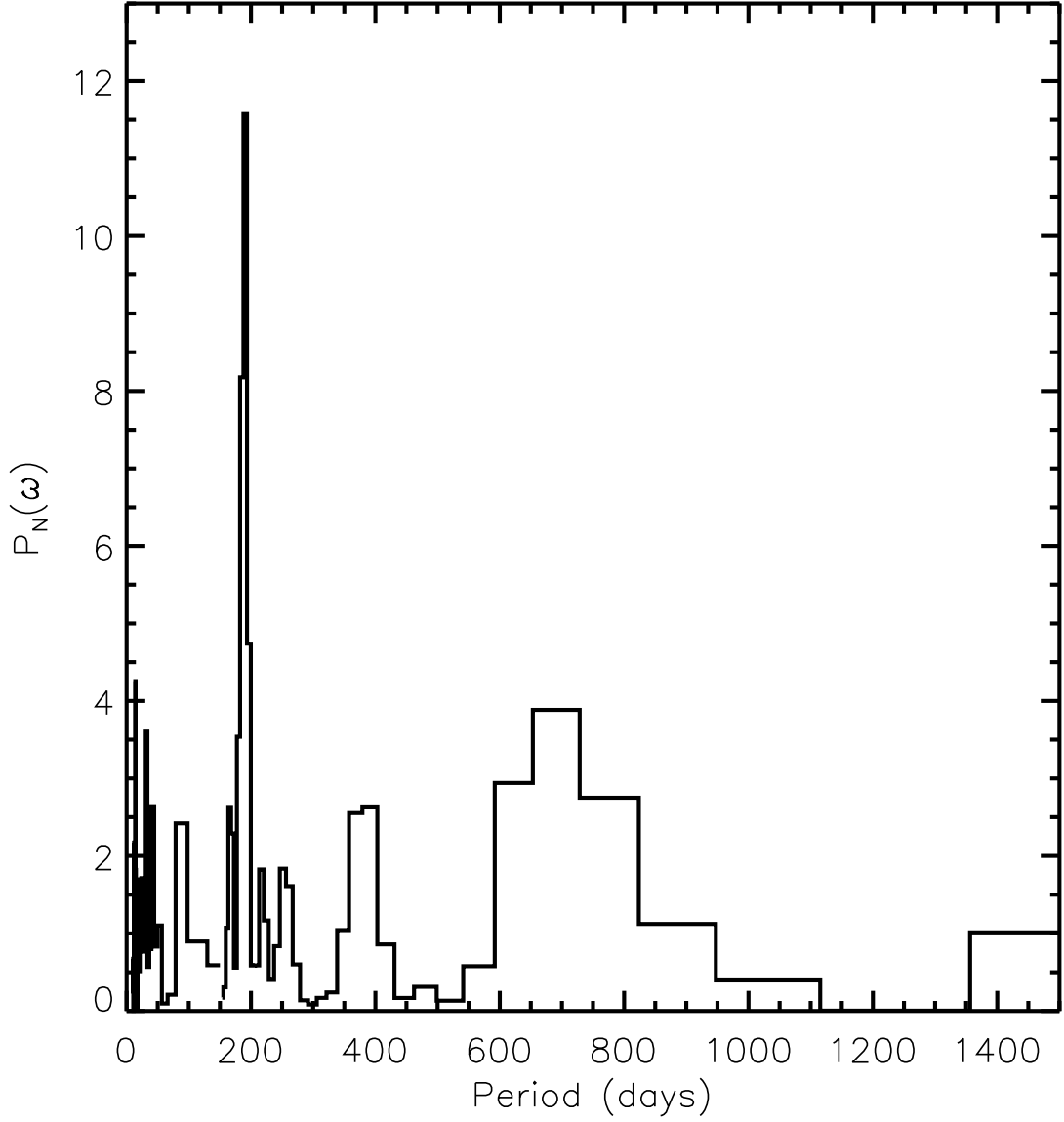}
\caption{A periodogram analysis of the X-ray light curve. The most significant period is $189 \pm 6$ days. Subsequent peaks appear at integer multiples of this period.}
\label{fig8}
\end{figure*}

\begin{figure*}
\plotone{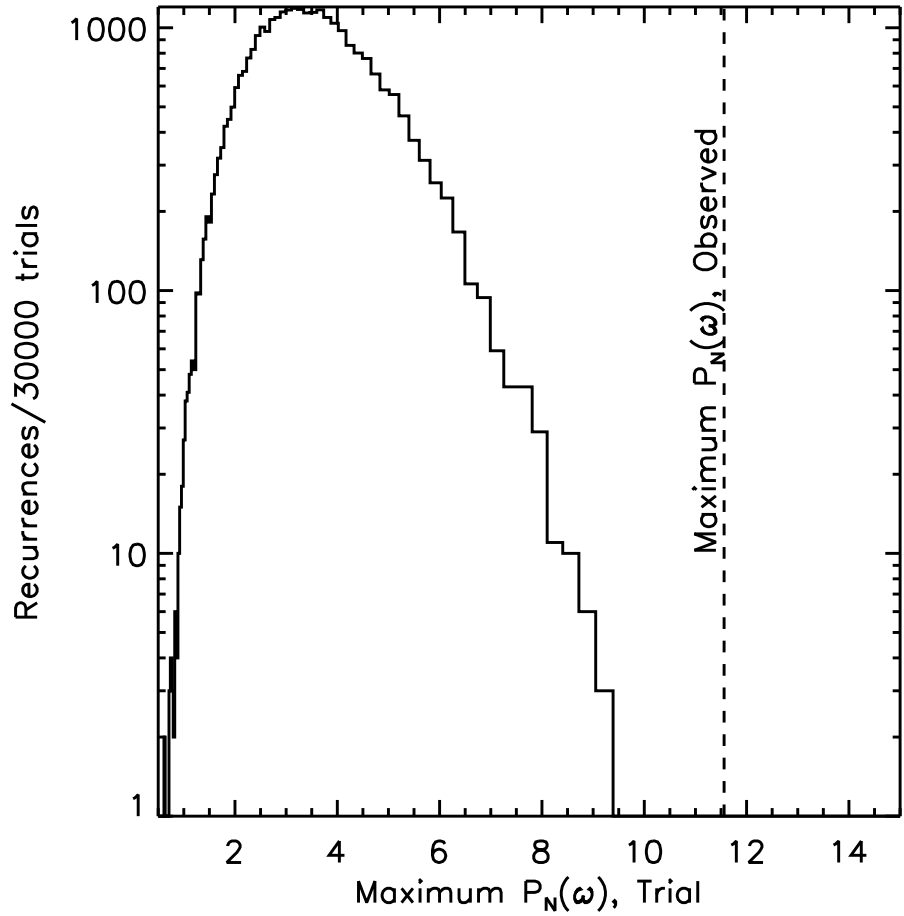}
\caption{Monte Carlo simulation testing the possibility of a random periodogram peak of the observed power (see Fig. \ref{fig8}) at this sampling. The vertical line indicates the power of the original signal. We find that our period is significant with a confidence level of 99.997\%.}
\label{fig9}
\end{figure*}

Because the X-ray flux is relatively constant over the course of each {\it XMM} observation, we calculate a single flux value for each observation epoch and combine these measurements with the {\it Chandra} light curve in Figure \ref{fig7}. Using the combined light curve, we can test for the presence or absence of periodic flux variations on timescales longer than 40 hours. The most notable flux variation in the {\it XMM} is a 4$\sigma$ variation in consecutive observations separated by five months (see Fig. \ref{fig6}). If periodic, this suggests a longer timescale variability. Using the method of \citet{horne86}, we perform a periodogram analysis of the combined {\it Chandra} and {\it XMM} light curve and find a period of $189 \pm 6$ days. In Figure \ref{fig8}, we show the resultant periodogram which tests for periodicity on scales of 1-1500 days. The peak at 189~d is clearly distinct and additional peaks are visible at integer multiples of the period. In Figure \ref{fig7}, we plot the X-ray light curve folded on the 189 d period. Analytically estimating the significance of a signal in non-uniformly sampled data is non-trivial. Thus, in order to estimate the confidence of this detection, we perform a Monte Carlo simulation as follows. We take the existing data set and maintain the same sampling intervals throughout. For each Monte Carlo realization, we randomly reassign the observed flux  values to the time samples, effectively scrambling the light curve. We plot the results of these simulations in Figure \ref{fig9}. In 30,000 trials, we do not achieve a peak power approaching the power of our original periodogram, implying that the 189-day period is not due to random noise with a confidence level greater than 99.997\%.

\begin{figure*}
\plotone{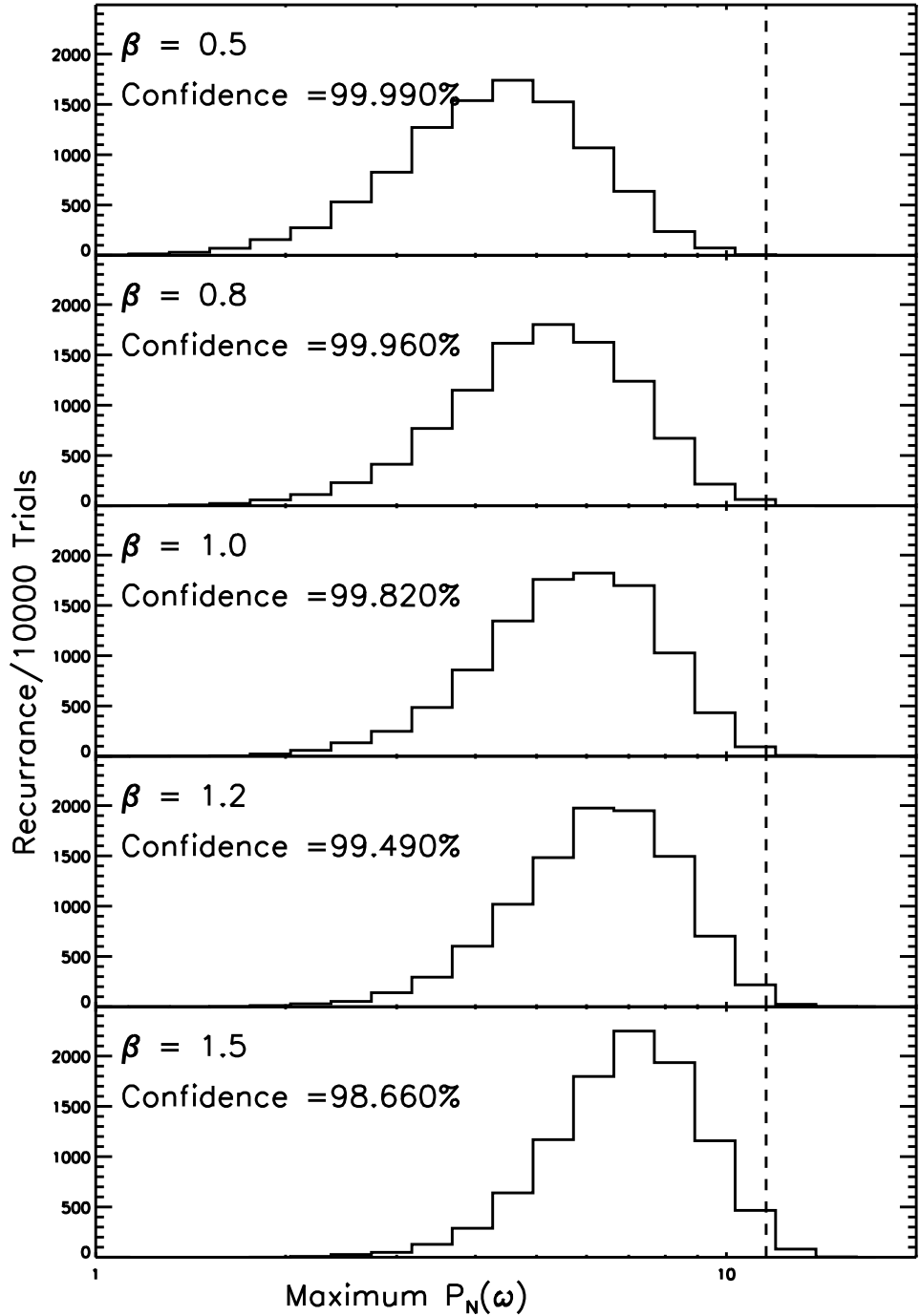}
\caption{Monte Carlo simulations testing for power peaks, as in Figure \ref{fig9}, but assuming different levels of red noise in the system (see text). As more red noise is assumed in the observation, the strength of the signal decreases.}
\label{fig10}
\end{figure*}

The previous test accounts for white noise variability; however, red noise is a significant source of false peaks in X-ray power spectra of X-ray binaries \citep{tit07}. Red noise is a flux variation in the power spectrum that can be parameterized with a frequency dependence $f^{-\beta}$. A white noise process will generate a flat power spectrum such that $\beta \sim 0$; a value of $\beta \sim 2$ describes random walk noise \citep{tk95}. A $\beta \sim 1$ dependence has been identified in stellar-mass black hole candidates and may be strongly related to accretion physics in the system \citep{min94,tk95,tit07}.  Following the method of \citet{tk95}, we test the possibility of red noise creating a false signal matching the strength of our periodogram. Simulating a number of red noise dominated light curves of varying power law slope, $\beta$, we find that as $\beta$ increases, more noise gets shunted near the period frequency, and the significance of our detection decreases. We show the results of our tests in Figure \ref{fig10}. We find the significance of our period detection remains above $3\sigma$ for values of $\beta \leq 1.0$ and above $2.5\sigma$ for $\beta \leq 1.5$, showing that the significance decreases slowly as red noise is increased.

\section{Discussion}

\subsection{Infrared Variability}

We test the IR spectra for variations on short timescales (hours to days).  Due to limits of our spectral resolution, we cannot observe radial velocity variations if the orbital velocity is less than 70km/s. In Figure \ref{fig4}, we show several close-ups of the Brackett-$\gamma$ region of CXOGC~J174536.1-285638's spectrum over the course of our three night IRTF run in 2006. To the left of the $\lambda$2.164$\mu$m marker, we see minor variances in the Helium contribution to the line. Because this line cannot be resolved from the larger Br-$\gamma$ contribution, it is difficult to determine the significance of this change. The RMS spectral difference rarely reaches 5$\sigma$ between any two events which are separated by $\sim 1$ hour. The observed differences are primarily in the wings of the line ($\sim 2.164 \mu$m or $\sim 2.168 \mu$m). Higher resolution spectroscopy is required to determine whether the changes in the Helium contribution are intrinsic to CXOGC~J174536.1-285638 rather than an artifact of the data analysis. The observed variations do not have any detectable periodicity. It should also be noted that this region is affected by the data reduction process, as described in \S 2. Our 2006 IR spectra were obtained about two days after the {\it Chandra} observations on Day 2402 in the X-ray light curve (see Fig. \ref{fig7}), where the object is transitioning from an apparent low-flux state to a high-flux state. Since we have no IR data consistent with the lowest X-ray flux events, it is impossible to determine from these IR data if the apparent Helium variability at $\lambda 2.164\mu$m we observe is associated with this X-ray flux transition.

We also search for wind variations in the P Cygni profiles. In our initial discovery spectrum, we identified three HeII lines with P~Cygni profiles: 2.0379 $\mu$m, 2.1891 $\mu$m, and 2.3464 $\mu$m (Paper I). In our 2005 analysis, we estimated the P~Cygni velocity at $170 \pm 70$ km/s. We repeat our analysis on the 2006 data to search for variations and find the approximate velocity of the wind is $200 \pm 70$ km/s. The error is dominated by the spectral resolution. We find no evidence of changes in the P~Cygni profile or velocity over our three day observations. Also, the 2005 and 2006 spectra have consistent P~Cygni profiles and velocities.

Unfortunately, it was not until after completion of our IR observation campaign that we discovered the 189~d X-ray periodicity in the source. Thus we were not able to schedule our IR observations to sample different X-ray phases; as a result, both our 2005 and 2006 observations sample the same phase (indicated in Figure \ref{fig7}). The lack of IR radial velocity variations is consistent with the observations being at the same phase of a long period system. 

\subsection{X-ray Variability}

Long term {\it Chandra} observations of this source revealed repeated X-ray flux variations, prompting us to search for periodicity by combining {\it XMM} and {\it Chandra} data, and revealing a 189-d period. In Paper I, we argue that CXOGC~J174536.1-285638 contains at least one massive star based on the presence of P~Cygni profiles in the IR spectrum. Although we consider the possibility of both an isolated massive star or a massive star in a binary system in Paper I, here we favor a binary interpretation because X-ray variability similar to that seen in CXOGC~J174536.1-285638 is not observed in isolated massive stars \citep{cohen00}. In comparing CXOGC~J174536.1-285638 to other systems containing massive stars, we showed that the X-ray to IR luminosity ratio, $L_X/L_K \sim 10^{-4}$, is consistent with both colliding wind binary (CWB) and high-mass X-ray binary (HMXB) systems (Paper I).

In the standard models for CWBs, X-ray emission arises from the shock front of colliding winds in two massive stars \citep[see, e.g.][]{luo90,sana04,debeck06}. Observed variability is often attributed to phase-locked flux modulations due to the effect of variations in absorption along the line of sight and variations in X-ray emission as a function of orbital phase. In this situation, the X-ray periodicity reflects an orbital period. Alternatively, it is possible that stellar rotation or photospheric pulsation may also produce periodic X-ray modulations. Models of such behavior are often employed to explain the 84~d quasi-periodicity in $\eta$ Carinae \citep{dav98}. In these situations, the modulation of the X-ray flux is correlated to recurrent behavior affecting the wind emission, but not related to the orbital period.
 
However, in HMXBs, periodic X-ray flux changes can be the result of either orbital or superorbital motion. A superorbital periodicity is defined as any periodicity apparent in the periodogram that is greater than the orbital period. The predominant model for superorbital periodicity is that of a precessing warped accretion disk; however, long period variations may also be due to the precession of a compact object (not applicable to black hole systems), periodic modulation of the mass accretion rate, or the influence of a third body \citep{paul00,og01,clarkson03}. Superorbital variations divide into two broad observational classes. The first class is characterized by clear, stable X-ray variations of about $\sim$30 days, while the second class has longer, quasi-periodic variations ranging from $\sim 50-200$ days \citep{clarkson03}. The second class is considered quasi-periodic, because long term monitoring shows a broad power peak in the periodogram, often superposed on a red noise spectrum \citep[e.g., Cyg X-2;][]{paul00}. Cen~X-3, Cyg~X-1, and Vela X-1 are all high-mass binary systems showing both orbital and superorbital periods. They range in X-ray luminosity from $L_X (2-8keV) \sim 10^{33.3 - 37.7} erg s^{-1}$ \citep[][and references therein]{mikles06}. The superorbital periods of these systems are 140 d, 142 d, and 93 d respectively and their orbital periods are 2.1 d, 5.6 d, and 8.9 d \citep{og01}. \citet{sood07} interprets these superorbital periods as unstable. Of the $\sim$20 sources for which both the orbital and superorbital period are known, no definitive empirical trend defines the relationship \citep[see Fig.1 of][]{sood07}. 

 The morphology of CXOGC~J174536.1-285638's X-ray light-curve is not inconsistent with that caused by a precessing accretion disk, in that the flux appears to vary uniformly in the hard and soft X-rays. However, there is presently no direct observational test to confirm that a period is superorbital rather than orbital. In order to verify the presence and physical cause of a superorbital period, additional physical parameters of the system are required, including the mass ratio of the system, the inclination of the disk with respect to the orbital plane, the orbital period, and the orbital separation \citep{clarkson03}. Thus, while we cannot rule out the possibility that this periodicity is superorbital, as yet, we do not have sufficient information to place meaningful constraints on the superorbital hypothesis. Thus for the remainder of this discussion, we restrict ourselves to exploring the possibility that the 189~d period is orbital rather than superorbital.

\subsection{The Orbital Period Assumption}
 For both the CWB and HMXB cases, the X-ray periodicity can trace the orbital period. CWBs have periods of days to years while HMXBs have shorter periods ranging from hours to days \citep{van98,mc03}. 
 In Paper I, we determine an absolute IR magnitude $M_K = -7.6 \pm 0.3$~mag for CXOGC~J174536.1-285638 using a distance of 8~kpc, reddening of $A_K =3.4$, and a 2MASS magnitude of $K_S =10.33$~mag. Given that the source appears blended in 2MASS, we verify the magnitude using the UKIDSS Galactic Plane Survey where the source is clearly resolved \citep{lawrence07,lucas08}. The UKIDSS survey lists the magnitude as $K = 10.390 \pm 0.001$~mag, which is consistent with 2MASS, given the photometric transform between the relevant filters in these two surveys is $<0.1$ mag. 

 We can use CXOGC~J174536.1-285638's exceptional brightness and the X-ray period to place constraints on the nature of the system. For our purposes, the ``primary'' star (mass, $M_{OB}$) will refer to the massive OB-star and the ``secondary'' star (mass, $M_2$) will refer to the companion whose nature has yet to be identified. 

\begin{figure*}
\includegraphics[angle=90, width=\textwidth]{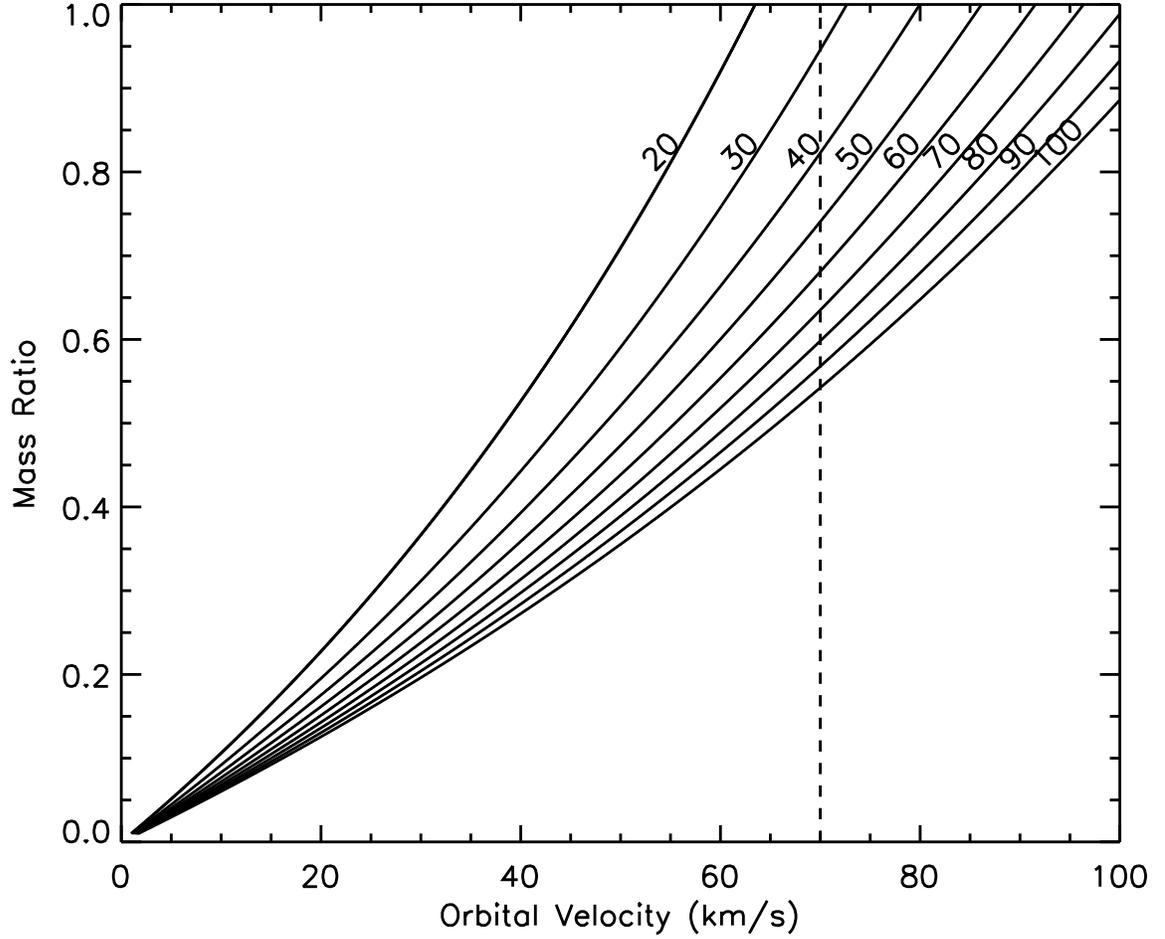}
\caption{Using the mass function and the putative period of 189 days, we calculate the expected mass ratio, $q=M_2/M_{OB}$, for primary masses $M_{OB} = 20 - 100 M_\odot$. The primary mass is indicated to the left of each line. The vertical dashed line represents the limiting IR spectral resolution.}
\label{fig11}
\end{figure*}

Using the mass function
\begin{displaymath}
f(q,i)=\frac{(q \sin{i})^3}{(1+q)^2} = \frac{P v_{orb}^3}{2 \pi G M_{OB}}
\end{displaymath}
where $q=M_2/M_{OB}$, we can generate a parameter space of orbital velocities and mass ratios for the system. Massive OB stars can range from $20 - 100 M_\odot$ and still emit strongly in the IR \citep[see, e.g.][]{cox00,girardi02}. 
In Figure \ref{fig11}, we plot the mass ratio as a function of the inferred orbital velocity for a range of primary masses and note that the orbital velocity is less than our IR spectral resolution of $70$~$km/s$ for cases of mass ratio $q<0.5$. Even for higher mass ratios, a radial velocity variation would have low signal-to-noise with our current observations. Thus, we require higher resolution spectroscopy in order to observe radial velocity variations in the IR associated with this periodicity.

In the next two sections we discuss the possibility that the modulations in X-ray flux are caused by (1) obscuration of the X-ray source by stellar wind; and (2) eclipse of the X-ray source. 

\subsubsection{Wind Obscuration Scenario}

Wind obscuration resulting in variable column absorption may be responsible, in part, for the X-ray flux variations observed in CXOGC~J174536.1-285638. This assumption would be most practically tested by analyzing the change in hardness, as softer X-ray photons are absorbed preferentially. Such analysis is hindered by the relative faintness of the X-ray source, i.e., the low count rate. For the spectra shown in Figure \ref{fig6}, the total integration time for each observation is 40 hours. For the higher flux observation on August 31, 2004, we observe a hardness ratio of $0.11 \pm 0.06$, where the soft counts are summed from $2-4$~keV, the hard counts $4-8$~keV. The hardness ratio is $(S-H)/(S+H)$ and the error is estimated from Poisson noise. For the second spectrum at the lower flux stage, taken on March 30, 2004, the hardness ratio is $0.01 \pm0.09$. The errors of these two measurements make them consistent with no change in hardness. However, the low count rate makes it difficult to estimate the robustness of this result. 

Energy-independent X-ray variations in the spectrum could result if electron scattering is an important source of absorption. By testing the possibility that an obscuring wind is solely responsible for the flux variations, we can find the upper limit of the mass loss rate of the massive star component of the system. If wind obscuration is only partially responsible for the flux variation, a lower mass-loss rate results. Thus, here, we are determining the most extreme wind-producing source required to produce the flux variations we observe.
 
 CXOGC~J174536.1-285638's X-ray light-curve shows a maximum flux variation by a factor of 4 over the course of the 189~d period. Using this information, if we assume that the X-ray emitting source is being obscured by a windy counterpart, we can calculate the column density of the wind required to cause such absorption. Because there are insufficient counts in the low-flux state to fit the X-ray spectrum, we use the model fit from the high-flux state and create a dummy response with XSPEC to measure the amount of absorption required to decrease the flux by a factor of four. Given our initial $N_H = 5.2 \times 10^{22} cm^{-2}$ (see Paper I), we find the column density from the obscuring wind must reach $N_H \approx 2.5 \times 10^{23} cm^{-2}$ to cause the flux variation observed in CXOGC~J174536.1-285638. 

To estimate the absorption column caused by a dense stellar wind, we use the equation:
\begin{equation}
N_H = \int_R^\infty \rho(r) dl.
\end{equation}
For a spherically symmetric shell, and a star with mass-loss rate $\dot{M}$ and escape velocity $V_\infty$,
\begin{equation}
\rho(r) = \frac{\dot{M}}{4\pi r^2 v_\infty}.
\end{equation}
For an edge-on view of the system, $dl = dr$, thus
\begin{equation}
N_H = \int_R^\infty \frac{\dot{M}}{4\pi v_\infty}\frac{dr}{r^2} = \frac{\dot{M}}{4\pi v_\infty R_{OB}}.
\end{equation}
Normalizing for typical values of $v_\infty =1000km/s$ and $\dot{M} =10^{-6}M_\odot yr^{-1}$ \citep[see, e.g.][]{mok07}, this becomes
\begin{equation}
\label{eq:nhmdot}
\frac{N_H}{10^{23} cm^{-2}} = 4.3 \times \left(\frac{\dot{M}}{10^{-6}M_\odot yr^{-1}}\right) \left(\frac{v_\infty}{1000km s^{-1}}\right)^{-1} \left(\frac{R_{OB}}{R_\odot}\right)^{-1}.
\end{equation}
If we are not viewing the system edge-on, we must take into account the angle through which we are viewing the wind as an effect on the observed absorption column. We can parameterize this in terms of an impact factor $b$ such that $b=r\cos{\theta}$. In this case, $dl = bd\theta$ and 
\begin{equation}
N_H = \frac{\dot{M}}{4 \pi v_\infty b} \int_{\theta_0}^{\pi/2} \cos^2{\theta}d\theta = \frac{\dot{M}}{4 \pi v_\infty b} \left[\frac{\pi}{2} - arccos\frac{b}{R} -\frac{b\sqrt{R^2-b^2}}{R^2}\right],
\end{equation}
where $\cos{\theta_0} = b/R$. Larger impact values require windier stars to create the same absorption column, thus the values of $\dot{M}$ estimated with Equation \ref{eq:nhmdot} should be considered a lower limit of the $\dot{M}$ required to produce the absorption column that causes the flux change in CXOGC~J174536.1-285638.

We estimate the mass loss for two special cases. In the first case, we postulate the IR light is dominated by a single bright source. In HMXBs, the star is expected to contribute more heavily to optical and IR emission than the accretion disk \citep{lewin97}. In certain CWB cases, especially of lower mass ratios, it is possible that a single source dominates emission \citep{lepine}. Thus for CWB and HMXB scenarios in which a single star dominates the IR emission, we use CXOGC~J174536.1-285638's IR luminosity and estimate stellar characteristics based on the isochrones of \citet{girardi02} and find that a star with $M_{K} \sim -7.6$ will likely have a radius $R_{OB} \sim 80 R_\odot$ valid for a range of masses $20-100 M_\odot$. Using equation \ref{eq:nhmdot}, we get a mass-loss rate of $\dot{M} \sim 4 \times 10^{-5} M_\odot yr^{-1}$. In the second case, we consider a system that contains two massive stars, each contributing half of the IR luminosity which is only consistent for CWBs containing two stars of similar bolometric luminosity. These stars would have $R_{OB} \sim 20R_\odot$ and $\dot{M} \sim 1 \times 10^{-5} M_\odot yr^{-1}$. Typical massive O-stars are reported to have mass-loss rates of $10^{-6} - 10^{-5} M_\odot yr^{-1}$ \citep{mok07}. Thus, even in the most extreme case, where the flux variation is caused entirely by absorption, a relatively windy star is necessary to produce the flux variations that we observe, but the mass-loss rate is not unreasonable.

\subsubsection{The Eclipsing Binary Scenario}

Assuming that the X-ray variability is caused by an eclipse has the greatest potential for constraining the nature of the system components, and also involves the most stringent physical constraints. We note that the X-ray light curve (Fig. \ref{fig7}) is atypical for a standard eclipsing source, both in the morphology of the dip and the phase duration of the low flux state. In a HMXB or CWB, the X-ray emitting region is small compared to the massive star. For a binary system in circular orbit, the eclipse of the X-ray region causes a decrease in X-ray emission that is relatively brief compared to the orbital period. For a binary system in an elliptical orbit, it is likely that the X-ray emitting region would experience periodic enhancement while the sources are in close approach. Our source spends approximately equal time at the high flux and low flux stage and transitions smoothly between the two. Despite this, we find it useful to explore the eclipsing assumption, as it allows us to define the limits of system in which the variation is caused by a combination of multiple effects (e.g., an eclipse plus wind obscurration).

By assuming that the low-flux portion of the dip is caused by an eclipse of the X-ray region, we estimate a transit time of $\tau \sim 50$ d for the putative eclipse, limited by adjacent observations of the high-flux stage. We convert the transit time to a velocity by estimating $v_{orb} = 2 R_{OB}/ \tau$. Combining this with the mass function, we get
\begin{equation}
\label{eq:massfunction}
\frac{(q \sin{i})^3}{(1+q)^2} = \frac{4 P R_{OB}^3}{\pi G M_{OB} \tau^{3}} = 6.5\times 10^{-7} \frac{(P/189d)}{(\tau/50d)^3}\frac{r_{OB}^3}{m_{OB}},
\end{equation}
where $m_{OB}$ and $r_{OB}$ are in units of solar masses and solar radii respectively. Assuming $\sin{i} =1$, we then solve the cubic equation for different scenarios. In Table \ref{tbl-4}, we list a series of mass ratios, $q$, associated with varying fractions $r_{OB}^3/m_{OB}$. As an example, we can examine the two cases as we did above. To complete this numerical exercise, we choose a median primary mass $M_{OB} =40M_\odot$ (while acknowledging that a wide range of masses is possible). 
If two massive stars each contribute half of the IR luminosity, then $R_{OB} \sim 20R_\odot$, $r_{OB}^3/m_{OB} = 200$, and the mass ratio is $q \sim 0.05$. This resulting mass ratio is inconsistent with our initial assumption of two massive stars contributing equally to the emission. If a single massive star dominates the IR emission, then $R_{OB} \sim 80 R_\odot$, $r_{OB}^3/m_{OB} = 12800$, and the mass ratio is $q \sim 0.2$. We find that adjusting the inclination does not significantly alter this result because ``eclipsing'' scenarios do not exist at low inclinations \citep[$i<82^o$;][]{terrell05}.

\begin{figure*}
\includegraphics[angle=90, width=\textwidth]{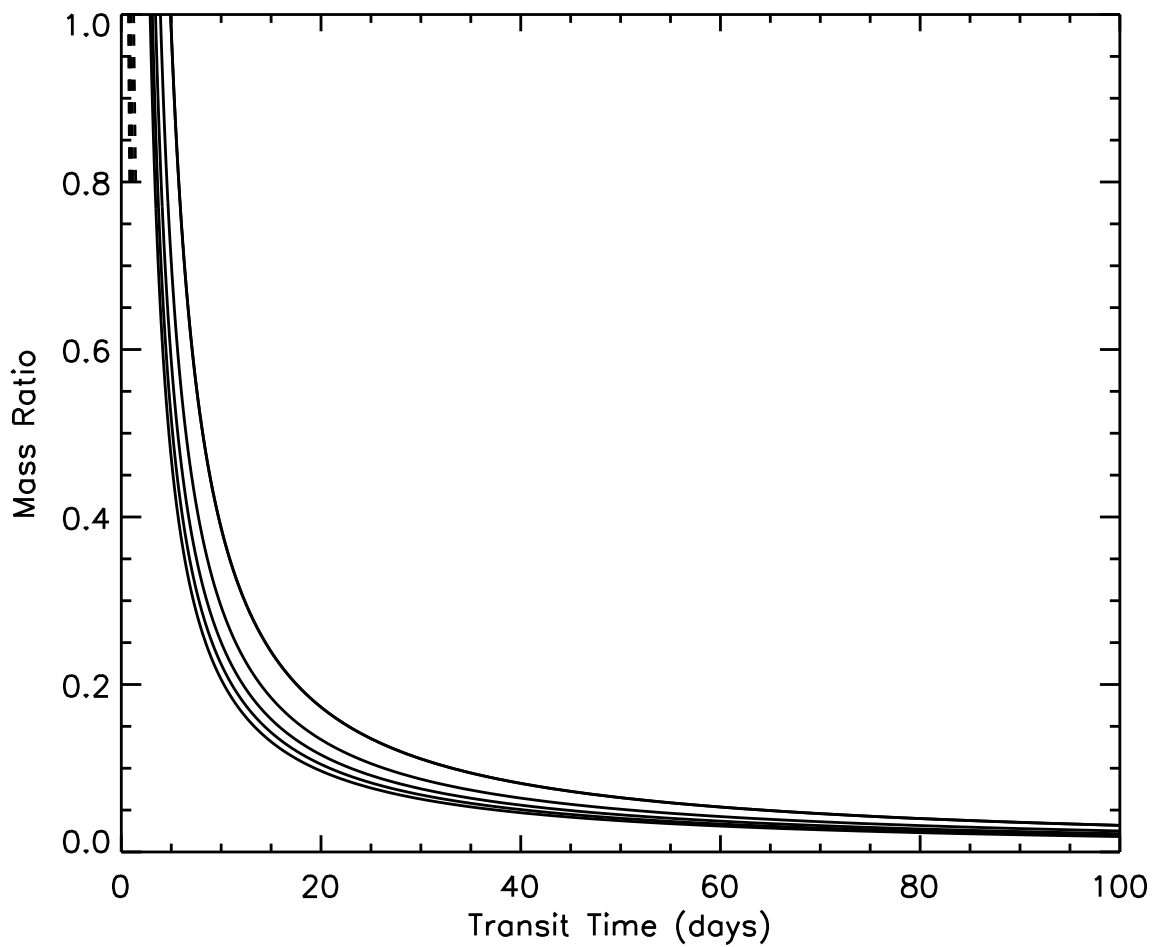}
\caption{In a similar manner to Figure \ref{fig11}, we compute the mass ratio of the system for a variety of transit times related to the orbital velocity of the system for primary sources ranging from $M_{OB} = 20 - 100 M_\odot$. The solid lines indicate systems in which a single massive source dominates the IR emission ($R = 80 R_\sun$) and the dashed line is for two massive sources contributing approximately equally to the emission ($R = 20 R_\sun$). See details in text. }
\label{fig12}
\end{figure*}

In Figure \ref{fig12}, we plot the mass ratio as a function of transit time, to explore the possibility that only a fraction of the flux variation is caused by an eclipse of the X-ray source. We convert the transit time into an orbital velocity using the radii $20 R_\odot$ and $80 R_\odot$ as we did above to represent systems where two massive stars contribute to the IR luminosity and systems where a single source dominates the IR emission. We find that for transit times above $\sim$10 days ($v_{orb} < 130 km/s$), the system is consistent with low mass ratios ($q < 0.4$). In systems where two stars are contributing equally to the IR luminosity (valid only for CWBs), the transit time would be $< 2$ days, corresponding to an orbital velocity $v_{orb} > 160$ days. Variability of this nature and on this timescale should have been apparent in our IR observations. Since we do not see those variations, we find eclipsing scenarios more likely for systems with lower mass ratios.

Thus if the system is a CWB, it would have to have a relatively low mass ratio with the IR emission dominated by a single source. This implies that the wind emission of one source overwhelms that of its companion \citep{luo90}. It is possible for CWBs to have lower mass ratios if the secondary is a Wolf-Rayet (WR) star. By the time a massive star reaches the WR stage, it may have a relatively small mass, but still have enormously powerful winds \citep{crowther07}. For example, $\gamma ^2 Velorum$ is a WR+O star with a mass ratio $q \sim 0.35$ \citep{vanderhutch01}. In the case of $\gamma ^2 Velorum$, the WR star dominates the IR emission, so the source appears Helium rich \citep{crowther07}. It is possible that the Helium emission we observe in CXOGC~J174536.1-285638 is evidence of an obscured WR companion. However, because Brackett series emission rather than Helium emission dominates the IR spectrum, we find this scenario less likely. In Table \ref{tbl-5}, we list line ratios of Br-$\gamma$ to HeI 2.114$\mu$m and Br-$\gamma$ to HeII 2.189$\mu$m for known CWBs and XRBs. In known WR+O binaries, the HeII 2.189$\mu$m is notably stronger than Br-$\gamma$. Comparatively, CXOGC~J174536.1-285638 has much stronger Br-$\gamma$ emission, and hence a quite different Br-$\gamma$/ HeII line ratio from what is observed in WR+O systems. In fact, we note the Br-$\gamma$/HeI and Br-$\gamma$/HeII line ratios in CXOGC~J174536.1-285638 are more consistent with HMXBs than either O+O or O+WR CWBs. Thus if CXOGC~J174536.1-285638 is a WR+O CWB, it is very unusual. In the eclipsing binary scenario, CXOGC~J174536.1-285638 would more likely be an HMXB.

\subsubsection{CXOGC~J174536.1-285638 as a Wind-Accreting HMXB}

In Paper I, we showed that the X-ray luminosity of CXOGC~J174536.1-285638 ($1.1 \times 10^{35}$ erg~s$^{-1}$) is consistent with HMXBs, within the observed range of X-ray luminosities between INTEGRAL sources identified as HMXBs \citep[$\sim 10^{34}$ erg~s$^{-1}$;][]{tomsick06, sidoli06} and the canonically bright sources such as Cyg~X-1 and Cen~X-3 \citep[$\sim 10^{37}$ erg~s$^{-1}$;][]{nag92,schulz02}. We explore the implications of the observed period for the case where CXOGC~J174536.1-285638 is an accreting binary system with a compact object.  Since the IR data suggest that CXOGC~J174536.1-285638 contains a high-mass star, we focus on the case of wind-fed accretion. 

Taking the standard accretion luminosity as 
\begin{equation}
L_X = \epsilon \dot{M} c^2,
\end{equation}
where $\epsilon$ is the efficiency of converting energy into X-ray light and $\dot{M}$ is the accretion rate, we can rewrite this in terms of the mass loss rate of the donor star due to wind such that

\begin{equation}
L_X \approx 5.7\times 10^{37} \epsilon \left(\frac{\dot{M}}{-10^{-4}\dot{M}_{wind}}\right)\left(\frac{-\dot{M}_{wind}}{10^{-5}M_\odot yr^{-1}}\right)erg~s^{-1}.
\end{equation}

We have normalized the mass loss rate of the primary due to wind and the accretion efficiency of the system with typical values found in \citet{fkr02}. \citet{fkr02} estimate the accretion efficiency, $\dot{M}/ \dot{M}_{wind}$, by comparing the mass flux within an accretion cylinder to the total mass loss of the donor star. The accretion cylinder is estimated from the gravitational potential of the compact object, giving
\begin{displaymath}
\frac{\dot{M}}{-\dot{M}_{wind}} = \frac{\pi r_{acc}^2 v_{wind}(a)}{4 \pi a^2 v_{wind}(a)}
\end{displaymath}
where $r_{acc} \sim 2 G M_2 / v_{wind}^2$, $v_{wind} \sim (2 G M_1 /R_1)^{1/2}$, and $a$ is the orbital separation. This gives us
 \begin{equation}
\frac{\dot{M}}{-\dot{M}_{wind}} \simeq \frac{1}{4}\left(\frac{M_{2}}{M_{OB}}\right)^2\left(\frac{R_{OB}}{a}\right)^2.
\end{equation}

Normalizing to standard values, and using our known values, we get
\begin{equation}
\label{eq:qmdotlx}
\frac{L_X}{10^{35} erg~s^{-1}} \approx 35 \epsilon \left[\frac{r_{OB}^3}{m_{OB}} \frac{q^3}{(1+q)}\right] \left[\frac{-\dot{M}_{wind}}{10^{-5}M_\odot yr^{-1}}\right]\left[\frac{P}{189~d}\right],
\end{equation}
where $r_{OB}$ and $m_{OB}$ are normalized to solar radii and solar masses respectively. This form is useful for exploring the scenarios put forth in the previous sections. Because we are considering a wind-accreting HMXB, we use our previous estimate where a single massive star dominates the system, for mass between $20 - 100 M_\odot$ and radius $R \sim 80 R_\odot$.

The wind obscuration scenario gave an estimate of $\dot{M}_{wind} \approx 4 \times 10^{-5} M_\odot/yr$. We can then use Equation \ref{eq:qmdotlx} and find that the mass ratio of the system is $q \sim 0.01$. This suggests a massive $M > 80 M_\odot$ donor for a typical neutron star companion. By relaxing the estimate of the massive star radius, $R_{OB}$, we find that $q$ will increase and more compact object solutions exist over a wider range of primary masses. The estimate of $R_{OB} = 80 R_\odot$ is derived from the gravitational potential as estimated by \citet{girardi02}. In Table \ref{tbl-6}, we list a series of solutions for Equation \ref{eq:qmdotlx}.

Because the eclipsing scenario case places firm constraints on the mass ratio of the system, we use Equation \ref{eq:qmdotlx} to calculate the mass loss rates associated with various scenarios. We list those values in Table \ref{tbl-4}. For the case where the mass ratio is $q \sim 0.2$, the associated mass loss rate is low ($\dot{M}_{wind} \sim 2 \times 10^{-7} M_\odot /yr$), for an efficiency $\epsilon \sim 0.1$. This is not unreasonable for massive stars \citep{mok07}. Interestingly, in both the wind obscuration and the eclipsing binary scenario, the X-ray luminosity is consistent with a low mass ratio for the system.

\section{Conclusions}

We have searched for evidence of periodic variability in the IR spectra and long-term X-ray light-curve of the GC X-ray source CXOGC~J174536.1-285638. We find no evidence of IR variability on short ($<3~d$) timescales or between the 2005 and 2006 spectra. We compare the IR line ratios Br-$\gamma$/HeI and Br-$\gamma$/HeII in CXOGC~J174536.1-285638 to known HMXBs and CWBs and find the relative emission line strengths to be more consistent with an HMXB. We have identified an apparent $189 \pm 6$ d period in the CXOGC~J174536.1-285638 X-ray light curve. We find no evidence of periodic X-ray variability at timescales less than 189~d. Using a Monte Carlo simulation, we test the significance of the 189~d period detection; despite our fairly sparse time sampling, we find this period is significant with a confidence level greater than 99.997\%. We explore several interpretations of the X-ray modulation.

It is plausible, if the source is a HMXB, that the periodic modulation is superorbital in nature and related to a precessing accretion disk, in which case, further observations are required to determine the orbital period of the system and thus the nature of the system components. If the source is a HMXB and the 189~d period is superorbital, then we expect to find a shorter orbital period. This putative orbital periodicity is not necessarily observable in the IR as in this scenario, the IR emission is dominated by a single bright source. If the orbital period is detectable in the X-ray, targeted observations with a sensitive detector over a time interval of $\sim 1-2$ weeks during the high flux stage are required to ensure sufficient counts to test for variability.

We also explore an orbital period interpretation and summarize scenarios for this in Table \ref{tbl-7}. If the observed period is orbital in nature, and the X-ray modulation is caused by obscuration of the X-ray source due to a dense wind, then CXOGC~J174536.1-285638 is consistent with both CWB and HMXB interpretations. The further constraint of the X-ray luminosity is consistent with a massive ($M_{OB} > 80 M_\odot$) donor with a neutron star companion. If X-ray modulation is caused by an eclipse, the mass ratio is low and CXOGC~J174536.1-285638 is more consistent with an HMXB interpretation. If the 189~d period is orbital, we may be able to identify the source nature by obtaining long term photometric observations in the IR. Also, targeted IR follow-up spectroscopy to cover multiple phases of the source period will allow us to search for a relationship between the X-ray and IR variability in this system. In the low flux phase, additional IR spectroscopic line features (e.g., absorption, P~Cygni variation) may become apparent that can help us discern the nature of the stellar components.

Recently, \citet{hyodo08} reported the discovery of an early-type,
Galactic Center source which appears to have many characteristics in
common with CXOGC~J174536.1-285638. The source, CXOGC
J174645.3-281546, has an unusually strong FeXXV line ($\sim$1~keV), shows
X-ray variability of a factor of $\sim 2$ on a $\sim$year timescale, and
appears to have a high-mass star as its likely IR counterpart.  As in
CXOGC~J174536.1-285638, its X-ray to IR luminosity ratio is $\sim
10^{-4}$.  These intriguing similarities in X-ray spectral appearance,
variability timescale, and luminosity lead us to suggest that it would be
interesting in future observations to study this source in concert with
CXOGC~J174536.1-285638.  Although there are only two sources with
these properties known at present, it is possible that they could
ultimately define a new (sub)class of early-type Galactic sources with
strong FeXXV emission.

\acknowledgements
The authors make use of observations obtained with XMM-Newton, an ESA science mission with instruments and contributions directly funded by ESA Member States and NASA.

Authors are Visiting Astronomers at the Infrared Telescope Facility, which is operated by the University of Hawaii under Cooperative Agreement no. NCC 5-538 with the National Aeronautics and Space Administration, Office of Space Science, Planetary Astronomy Program. Many thanks to the IRTF support staff who assisted us with remote observing on this run. 

VJM, SSE, and RMB are supported in part by an NSF Grant (AST-0507547).
MPM was supported by the National Aeronautics and Space Administration through Chandra Award Number GO6-7135 issued by the Chandra X-ray Observatory Center, which is operated by the Smithsonian Astrophysical Observatory for and on behalf of the National Aeronautics Space Administration under contract NAS8-03060.

\clearpage

\begin{deluxetable}{llll}
\tablecaption{{\bf Observing Log: IR Spectra}}
\tablewidth{0pt}
\tablehead{\colhead{Obs ID} & \colhead{Date} & \colhead{Time (UT)} & \colhead{Exposure Time (min)}}
\startdata
a    &   2006-08-02    &     6:38 &  20  \\
b    &   2006-08-02    &     7:05 &  20   \\
c    &   2006-08-02    &     8:25 &  20   \\
d    &   2006-08-02    &     8:55 &  16   \\
e    &   2006-08-02    &     9:46 &   8   \\
f    &   2006-08-03    &     5:27 &  16   \\
g    &   2006-08-03    &     5:52 &  16   \\
h    &    2006-08-03   &     6:37 &  20   \\
i    &    2006-08-03   &     7:07 &  20  \\
j    &    2006-08-03   &     7:49 &  20  \\
k    &    2006-08-03   &     8:20 &  20  \\
l    &    2006-08-03   &     9:02 &  20  \\
m    &    2006-08-03   &     9:32 &  12  \\
n    &    2006-08-04   &     5:51 &  20  \\
o   &    2006-08-04   &     6:27 &  16  \\
p   &    2006-08-04   &     6:46 &  16  \\
q    &    2006-08-04   &     7:23 &  20  \\
r   &    2006-08-04   &     7:58 &  16  \\
s   &    2006-08-04   &     8:11 &  18  \\
t   &    2006-08-04   &     8:47 &  20  \\
u   &    2006-08-04   &     9:10 &  20  \\   				
\tableline			
\enddata			
\tablecomments{These observation IDs are associated with Figures 1 and 2. The days align with days 2404-2406 on our X-ray light curves.}
\label{tbl-1}
\end{deluxetable}

\begin{deluxetable}{ccccccc}
\tablecaption{{\bf Observing Log: Chandra }}
\tablewidth{0pt}
\tablecolumns{7}
\tablehead{Date & Time & Obs. ID & Exp. Time & R.A. & Declination & Roll \\
\multicolumn{2}{c}{(UT)} & & (ks) & \multicolumn{2}{c}{(J2000)} & (deg)}
\startdata
2000-10-26 & 18:15:11 & 1561a & 35.7 & 266.41344 & -29.01281 & 264.7\\
2001-07-14 & 01:51:10 & 1561b & 13.5 & 266.41344 & -29.01281 & 264.7\\
2001-07-18 & 14:25:48 & 2284  & 10.6 & 266.40415 & -28.94090 & 283.8\\
2002-05-22 & 22:59:15 & 2943  & 34.7 & 266.41991 & -29.00407 & 75.5\\
2002-02-19 & 14:27:32 & 2951  & 12.4 & 266.41867 & -29.00335 & 91.5\\
2002-03-23 & 12:25:04 & 2952  & 11.9 & 266.41897 & -29.00343 & 88.2\\
2002-04-19 & 10:39:01 & 2953  & 11.7 & 266.41923 & -29.00349 & 85.2\\
2002-05-07 & 09:25:07 & 2954  & 12.5 & 266.41938 & -29.00374 & 82.1\\
2002-05-25 & 15:16:03 & 3392  & 165.8 & 266.41992 & -29.00408 & 75.5\\
2002-05-28 & 05:34:44 & 3393  & 157.1 & 266.41992 & -29.00407 & 75.5\\
2003-06-19 & 18:28:55 & 3549  & 24.8 & 266.42092 & -29.01052 & 346.8\\
2002-05-24 & 11:50:13 & 3663  & 38.0 & 266.41993 & -29.00407 & 75.5\\
2002-06-03 & 01:24:37 & 3665  & 89.9 & 266.41992 & -29.00407 & 75.5\\
2004-07-05 & 22:33:11 & 4683  & 49.5 & 266.41605 & -29.01238 & 286.2\\
2004-07-06 & 22:29:57 & 4684  & 49.5 & 266.41597 & -29.01236 & 285.4\\
2004-08-28 & 12:03:59 & 5360  &  5.1 & 266.41477 & -29.01211 & 271.0\\
2005-07-24 & 19:58:27 & 5950  & 48.5 & 266.41519 & -29.01222 & 276.7\\
2005-07-27 & 19:08:16 & 5951  & 44.6 & 266.41512 & -29.01219 & 276.0\\
2005-07-29 & 19:51:11 & 5952  & 43.1 & 266.41508 & -29.01219 & 275.5\\
2005-07-30 & 19:38:31 & 5953  & 45.4 & 266.41506 & -29.01218 & 275.3\\
2005-08-01 & 19:54:13 & 5954  & 18.1 & 266.41502 & -29.01215 & 274.9\\
2005-02-27 & 06:26:04 & 6113  &  4.9 & 266.41870 & -29.00353 & 90.6\\
2006-07-17 & 03:58:28 & 6363  & 29.8 & 266.41541 & -29.01228 & 279.5\\
2006-04-11 & 05:33:20 & 6639  &  4.5 & 266.41890 & -29.00369 & 86.2\\
2006-05-03 & 22:26:26 & 6640  &  5.1 & 266.41935 & -29.00383 & 82.8\\
2006-06-01 & 16:07:52 & 6641  &  5.1 & 266.42018 & -29.00440 & 69.7\\
2006-07-04 & 11:01:35 & 6642  &  5.1 & 266.41633 & -29.01237 & 288.4\\
2006-07-30 & 14:30:26 & 6643  &  5.0 & 266.41510 & -29.01218 & 275.4\\
2006-08-22 & 05:54:34 & 6644  &  5.0 & 266.41484 & -29.01202 & 271.7\\
2006-09-25 & 13:50:35 & 6645  &  5.1 & 266.41448 & -29.01195 & 268.3\\
2006-10-29 & 03:28:20 & 6646  &  5.1 & 266.41425 & -29.01178 & 264.4\\
\tableline
\enddata
\label{tbl-2}
\end{deluxetable}

\begin{deluxetable}{llll}
\tablecaption{{\bf Observing Log: XMM-Newton }}
\tablewidth{0pt}
\tablecolumns{11}
\tablehead{\colhead{Observation ID} & \colhead{Date}  & \colhead{Time (h)} & \colhead{Exposure Time (h)}}
\startdata
0112972101 & 2001-09-04 & 01:19:34 & 7.5 \\
0111350101 & 2002-02-26 & 03:11:27 & 14  \\
0111350301 & 2002-10-03 & 06:36:49 & 5   \\
0202670501 & 2004-03-28 & 14:37:16 & 40 \\
0202670601 & 2004-03-30 & 14:29:07 & 40 \\
0202670701 & 2004-08-31 & 02:54:31 & 40  \\
0202670801 & 2004-09-02 & 02:44:08 & 40 \\
\tableline
\enddata
\label{tbl-3}
\end{deluxetable}

\begin{deluxetable}{lll}
\tablecaption{{\bf Mass ratio estimations for the eclipsing scenario}}
\tablewidth{4in}
\tablehead{\colhead{$r_{OB}^3 / m_{OB}$} & \colhead{$M_2/M_{OB}$} & $M_{wind}/ M_\odot~yr^{-1}$ \\}
\startdata
$10^5$ & 0.5  & $8 \times 10^{-9}$ \\
$10^4$ & 0.2  & $2 \times 10^{-7}$ \\
$10^3$ & 0.09  & $4 \times 10^{-6}$ \\
$10^2$ & 0.04 & $1 \times 10^{-4}$  \\
$10^1$ & 0.02 & $3 \times 10^{-2}$  \\
\tableline
\enddata
\tablecomments{The mass ratio expected for a primary of the given mass to radius ratio in the eclipsing binary scenario. In Column 1, the ratios are in units of $R_\odot^3/M_\odot$. Values of $r_{OB}^3 / m_{OB} > 10^{4}$ are more typical of brighter stars ($M_K \sim -7.6$) and thus consistent with cases where a single massive star is dominating CXOGC~J174536.1-285638's IR emission. Values of $r_{OB}^3 / m_{OB} < 10^{4}$ are more consistent with $M_K \sim -4$ stars such that CXOGC~J174536.1-285638's IR emission is composed of the flux from two bright stars. The estimation of $M_{wind}$ is based on Equation \ref{eq:qmdotlx}, which is only valid for the HMXB case.}
\label{tbl-4}
\end{deluxetable}

\begin{deluxetable}{lllllll}
\tablecaption{{\bf Infrared Line Ratios}}
\tablewidth{\textwidth}
\scriptsize
\tablehead{Source & \multicolumn{3}{l}{Equivalent Width ($\AA$)} & Ref. & Br$\gamma$/HeI & Br$\gamma$/HeII\\
             & HeI & Br$\gamma$ & HeII & & & \\
             & 2.114$\mu$m & 2.166$\mu$m & 2.189$\mu$m & & & }
\startdata
CXOGC~J174536.1-285638 & 13.8 & 36.6 & $<$2 & 1 & 2.65 & $>$18.3 \\
\tableline
{\bf HMXB} & & & & & & \\
Cir X-1 & & 24.2 & 1.3 & 2 & & 18.62 \\
IGR J16318-4848 (sgB[e])  & 5 & 45 & & 5 & 9 & \\
HD 34921 (B0I) & 1 & 6 & & 6 & 6 & \\
HD 24534 (O9III-Ve) & 2.7 & 14.5 & & 2 & 5.37 & \\
EXO2030+375 & 1.7 & 4 & & 2 & 2.35 & \\
V725Tau (O9.7IIe) & & 13 & $<1$ & 6 & & $>13$ \\
\tableline
{\bf O+O} & & & & & & \\
HD 93205 (O3V)		& &	2    &	1.1    &      	6 &  & 1.82 \\
HD 206267 (O6.5V)	& &	1.2  &	0.4    &	6 &  & 3 \\
HD 152248 (O7Ib)	& &	4    &	1.8    &	6 &  & 2.22 \\
HD 57060 (O7Ia)		& &	5    &	1.1    &	6 &  & 4.55 \\
HD 47129 (O8)		& &	7    &	$<0.5$ &	6 &  & $>14$\\
HD 37043 (O9III)	& &	1.6  &	0.2    &	6 &  & 8\\
HD 47129(O7.5I+O6I)	& &	7    &	$<0.5$ &	6 &  &  $>14$\\
HD 15558 (O5III)	& &	1.4  &	0.4    &	6 &  & 3.5 \\
HD 199579 (O6V)		& &	1.4  &	0.6    &	6 &  & 2.33 \\
\tableline
{\bf O+WR} & & & & & & \\
WR138 (WN5+O9)	     & 12   &	34   &	52   &	4   &	2.83  &	0.65 \\
WR139 (WN5+O6)	     & 15   &	28   &	66   &	4   &	1.87  &	0.42\\
WR133 (WN4.5+O9.5)   &	    &	30   &	20   &	4   &	      &	0.63\\
WR127 (WN4+O9.5)     & 16   &	41   &	77   &	4   &	2.56  &	0.53\\
WR151 (WN4+O8)	     & 16   &	36   &	81   &	4   &	2.25  &	0.44\\
\tableline			
\enddata			
\tablecomments{IR line ratios. We compare the relative strength of HeI and HeII lines to Br-$\gamma$ in CXOGC~J174536.1-285638 and a selection of HMXBs and CWBs. Note that the HeII 2.189$\mu$m line in CXOGC~J174536.1-285638 has a P Cygni profile. We group O+O and O+WR binaries separately, as the former systems are less likely to produce low mass ratios. In known WR+O systems, the Br-$\gamma$/HeII line ratio is significantly different than that observed in CXOGC~J174536.1-285638. REFERENCES - (1) Paper 1; (2) \citet{clark99}; (3) \citet{clark03}; (4) \citet{figer97}; (5) \citet{fill04}; (6) \citet{hanson96}.}
\label{tbl-5}
\end{deluxetable}

\begin{deluxetable}{llll}
\tablecaption{{\bf Mass ratio estimations for the wind obscuration scenario in the case of a HMXB}}
\tablewidth{4in}
\tablehead{\colhead{$R_{OB}/ R_\odot$} & \colhead{$M_{OB}/M_\odot$} &  \colhead{$M_2/M_{OB}$} &  \colhead{$M_2/M_\odot$} }
\startdata
80  &  20  & 0.010 & 0.2 \\
\tableline
80  &  60  & 0.011 & 0.6 \\
\tableline
80  &  100  & 0.016 & 1.6 \\
\tableline\tableline
50  &  20  & 0.015 & 0.3 \\
\tableline
50  &  60  & 0.022 & 1.3 \\
\tableline
50  &  100  & 0.026 & 2.6 \\
\tableline\tableline
20  &  20  & 0.03 & 0.6 \\
\tableline
20  &  60  & 0.06 & 3.6 \\
\tableline
20  &  100  & 0.07 & 7.0 \\
\tableline
\enddata
\tablecomments{The mass ratio and compact object mass expected for a primary of the given mass to radius ratio in the wind obscuration scenario, valid for the HMXB case. The estimation of $q$ is based on Equation \ref{eq:qmdotlx}. We use $L_X = 1.1 \times 10^{35} erg~s^{-1}$ and assume an efficiency $\epsilon = 0.1$, and a mass loss rate $\dot{M} = 4 \times 10^{-5} M_\odot~yr^{-1}$. The value $R_{OB} = 80 R_\odot$ is most consistent with our observed IR luminosity \citep{girardi02}.}
\label{tbl-6}
\end{deluxetable}

\begin{deluxetable}{p{3.2in}|p{3.2in}}
\tablecaption{{\bf Summary of scenarios under the orbital period assumption}}
\tablewidth{\textwidth}
\tablehead{\colhead{\bf WIND OBSCURATION SCENARIO} & \colhead{\bf ECLIPSING BINARY SCENARIO}}
\startdata
\multicolumn{2}{c}{Two stars contributing equally to the IR luminosity (CWB)}\\
\multicolumn{2}{c}{$R_{OB} \sim 20 R_\odot$}\\
$\dot{M} = 10^{-5} M_\odot /yr$ (Eq. \ref{eq:nhmdot}) & $q \approx 0.05$ (Eq. \ref{eq:massfunction})\\
{\it consistent}  & {\it inconsistent with initial assumptions} \\
\tableline
\multicolumn{2}{c}{One star dominating the IR luminosity (CWB)}\\
\multicolumn{2}{c}{$R_{OB} \sim 80 R_\odot$}\\
$\dot{M} = 4\times 10^{-5} M_\odot /yr$ (Eq. \ref{eq:nhmdot}) & $q \approx 0.2$ (Eq. \ref{eq:massfunction}) \\
{\it consistent}  & {\it IR line ratios inconsistent with known WR+O systems} \\
\tableline
\multicolumn{2}{c}{One star dominating the IR luminosity (HMXB)}\\
\multicolumn{2}{c}{$R_{OB} \sim 80 R_\odot$}\\
\multicolumn{2}{c}{$L_X = 1.1 \times 10^{35} erg~ s^{-1}$}\\
$\dot{M} = 4\times 10^{-5} M_\odot /yr$ (Eq. \ref{eq:nhmdot}) & $q \approx 0.2$  (Eq. \ref{eq:massfunction})\\
$q \sim 0.01$   (Eq. \ref{eq:qmdotlx})    & $\dot{M} = 2 \times 10^{-7} M_\odot /yr$ (Eq. \ref{eq:qmdotlx})\\
{\it radius constraint suggests $M_{OB} > 80 M_\odot$}  & {\it consistent} \\

\enddata
\label{tbl-7}
\tablecomments{See details of more general cases and caveats in Section 3.3.3.}
\end{deluxetable}


\begin{thebibliography}{}

\bibitem[Beckmann et al.(2005)]{beck05} Beckmann, V., et al. \
2005, \apj, 631, 506 

\bibitem[Benaglia et al.(2001)]{ben01} Benaglia, P., Cappa, 
C.~E., \& Koribalski, B.~S.\ 2001, \aap, 372, 952 

\bibitem[Bodaghee et al.(2007)]{bod07} Bodaghee, A., et al. \
2007, \aap, 467, 585 

\bibitem[Clark \& Dolan(1999)]{clark99} Clark, L.~L., \& Dolan, 
J.~F.\ 1999, \aap, 350, 1085 

\bibitem[Clark et al.(2003)]{clark03} Clark, J.~S., Charles, 
P.~A., Clarkson, W.~I., \& Coe, M.~J.\ 2003, \aap, 400, 655

\bibitem[Clarkson et al.(2003)]{clarkson03} Clarkson, W.~I., 
Charles, P.~A., Coe, M.~J., Laycock, S., Tout, M.~D., 
\& Wilson, C.~A.\ 2003, \mnras, 339, 447 

\bibitem[Cohen(2000)]{cohen00} Cohen, D.~H.\ 2000, IAU 
Colloq.~175: The Be Phenomenon in Early-Type Stars, 214, 156 

\bibitem[Corbet(1986)]{corbet86} Corbet, R.~H.~D.\ 1986, \mnras, 
220, 1047

\bibitem[Cox(2000)]{cox00} Cox, A.~N.\ 2000, Allen's 
Astrophysical Quantities, 4th ed.~Publisher: New York: AIP Press; Springer, 
2000.~Edited by Arthur N.~Cox.~ ISBN: 0387987460

\bibitem[Crowther(2007)]{crowther07} Crowther, P.~A.\ 2007, \araa, 
45, 177 

\bibitem[Cushing et al.(2004)]{cushing04} Cushing, M.~C., Vacca, 
W.~D., \& Rayner, J.~T.\ 2004, \pasp, 116, 362 

\bibitem[Davidson et al.(1998)]{dav98} Davidson, K., 
Ishibashi, K., \& Corcoran, M.~F.\ 1998, New Astronomy, 3, 241 

\bibitem[De Becker et al.(2006)]{debeck06} De Becker, M., Rauw, 
G., Sana, H., Pollock, A.~M.~T., Pittard, J.~M., Blomme, R., Stevens, 
I.~R., \& van Loo, S.\ 2006, \mnras, 371, 1280 

\bibitem[Eikenberry et al.(2001)]{eiken01} Eikenberry, S.~S., 
Cameron, P.~B., Fierce, B.~W., Kull, D.~M., Dror, D.~H., Houck, J.~R., 
\& Margon, B.\ 2001, \apj, 561, 1027 

\bibitem[Fender et al.(2003)]{fender03} Fender, R., Migliari, 
S., \& M{\'e}ndez, M.\ 2003, New Astronomy Review, 47, 481 

\bibitem[Figer et al.(1997)]{figer97} Figer, D.~F., McLean, I.~S., \& Najarro, F.\ 1997, \apj, 486, 420 

\bibitem[Filliatre \& Chaty(2004)]{fill04} Filliatre, P., \& 
Chaty, S.\ 2004, \apj, 616, 469 

\bibitem[Frank et al.(2002)]{fkr02} Frank, J., King, A., \& 
Raine, D.~J.\ 2002, Accretion Power in Astrophysics, by Juhan Frank and 
Andrew King and Derek Raine, pp.~398.~ISBN 0521620538.~Cambridge, UK: 
Cambridge University Press, February 2002.

\bibitem[Girardi et al.(2002)]{girardi02} Girardi, L., Bertelli, 
G., Bressan, A., Chiosi, C., Groenewegen, M.~A.~T., Marigo, P., Salasnich, 
B., \& Weiss, A.\ 2002, \aap, 391, 195 

\bibitem[Hanson et al.(1996)]{hanson96} Hanson, M.~M., Conti, 
P.~S., \& Rieke, M.~J.\ 1996, \apjs, 107, 281 

\bibitem[Horne \& Baliunas(1986)]{horne86} Horne, J.~H., \& 
Baliunas, S.~L.\ 1986, \apj, 302, 757 

\bibitem[Hyodo et al.(2008)]{hyodo08} Hyodo, Y., Tsujimoto, M., 
Koyama, K., Nishiyama, S., Nagata, T., Sakon, I., Murakami, H., 
\& Matsumoto, H.\ 2008, \pasj, 60, 173 

\bibitem[Kirsch et al.(2005)]{kirsch05} Kirsch, M.~G., et al.\ 
2005, \procspie, 5898, 212 

\bibitem[Lawrence et al.(2007)]{lawrence07} Lawrence, A., et al.\ 
2007, \mnras, 379, 1599 

\bibitem[L{\'e}pine et al.(2001)]{lepine} L{\'e}pine, S., 
Wallace, D., Shara, M.~M., Moffat, A.~F.~J., \& Niemela, V.~S.\ 2001, \aj, 
122, 3407

\bibitem[Lewin et al.(1997)]{lewin97} Lewin, W.~H.~G., van 
Paradijs, J., \& van den Heuvel, E.~P.~J.\ 1997, X-ray Binaries, Edited by 
Walter H.~G.~Lewin and Jan van Paradijs and Edward P.~J.~van den Heuvel, 
pp.~674.~ISBN 0521599342.~Cambridge, UK: Cambridge University Press, 
January 1997.

\bibitem[Lewin \& van der Klis(2006)]{mc03} Lewin, W.~H.~G., 
\& van der Klis, M.\ 2006, Compact stellar X-ray sources.

\bibitem[Liu et al.(2006)]{liu06} Liu, Q.~Z., van Paradijs, 
J., \& van den Heuvel, E.~P.~J.\ 2006, \aap, 455, 1165 

\bibitem[Lucas et al.(2007)]{lucas08} Lucas, P.~W., et al.\ 
2007, ArXiv e-prints, 712, arXiv:0712.0100

\bibitem[Luo et al.(1990)]{luo90} Luo, D., McCray, R., \& Mac 
Low, M.-M.\ 1990, \apj, 362, 267

\bibitem[Mikles et al.(2006)]{mikles06} Mikles, V.~J., 
Eikenberry, S.~S., Muno, M.~P., Bandyopadhyay, R.~M., \& Patel, S \ 2006, 
\apj, 651, 408 

\bibitem[Mineshige et al.(1994)]{min94} Mineshige, S., Ouchi, 
N.~B., \& Nishimori, H.\ 1994, \pasj, 46, 97 

\bibitem[Mokiem et al.(2007)]{mok07} Mokiem, M.~R., et al.\ 
2007, \aap, 465, 1003 

\bibitem[Muno et al.(2003)]{mun03} Muno, M.~P., et al.\ 2003, 
\apj, 589, 225 

\bibitem[Muno et al.(2004a)]{mun04b} Muno, M.~P., et al.\ 2004a, 
\apj, 613, 326 

\bibitem[Muno et al.(2004b)]{mun04} Muno, M.~P., et al.\ 2004b, 
\apj, 613, 1179 

\bibitem[Muno et al.(2006)]{mun06} Muno, M.~P., Bauer, F.~E., 
Bandyopadhyay, R.~M., \& Wang, Q.~D.\ 2006, \apjs, 165, 173 

\bibitem[Muno et al.(2006)]{mun05} Muno, M.~P., Bower, G.~C., 
Burgasser, A.~J., Baganoff, F.~K., Morris, M.~R., 
\& Brandt, W.~N.\ 2006, \apj, 638, 183 

\bibitem[Nagase et al.(1992)]{nag92} Nagase, F., Corbet, 
R.~H.~D., Day, C.~S.~R., Inoue, H., Takeshima, T., Yoshida, K., \& Mihara, 
T.\ 1992, \apj, 396, 147 

\bibitem[Negueruela et al.(2005)]{neg05} Negueruela, I., 
Smith, D.~M., Reig, P., Chaty, S., \& Torrej{\'o}n, J.~M.\ 2006, The X-ray 
Universe 2005, 604, 165 

\bibitem[Negueruela \& Schurch(2007)]{neg07} Negueruela, I., 
\& Schurch, M.~P.~E.\ 2007, \aap, 461, 631 

\bibitem[Ogilvie \& Dubus(2001)]{og01} Ogilvie, G.~I., \& 
Dubus, G.\ 2001, \mnras, 320, 485

\bibitem[Paul et al.(2000)]{paul00} Paul, B., Kitamoto, S., 
\& Makino, F.\ 2000, \apj, 528, 410 

\bibitem[Predehl \& Schmitt(1995)]{ps95} Predehl, P., \& 
Schmitt, J.~H.~M.~M.\ 1995, \aap, 293, 889

\bibitem[Rayner et al.(2003)]{rayner03} Rayner, J.~T., Toomey, 
D.~W., Onaka, P.~M., Denault, A.~J., Stahlberger, W.~E., Vacca, W.~D., 
Cushing, M.~C., \& Wang, S.\ 2003, \pasp, 115, 362 

\bibitem[Sana et al.(2004)]{sana04} Sana, H., Stevens, I.~R., 
Gosset, E., Rauw, G., \& Vreux, J.-M.\ 2004, \mnras, 350, 809 

\bibitem[Schulz et al.(2002)]{schulz02} Schulz, N.~S., Cui, W., 
Canizares, C.~R., Marshall, H.~L., Lee, J.~C., Miller, J.~M., \& Lewin, 
W.~H.~G.\ 2002, \apj, 565, 1141 

\bibitem[Sidoli et al.(2006)]{sidoli06} Sidoli, L., Mereghetti, 
S., Favata, F., Oosterbroek, T., \& Parmar, A.~N.\ 2006, \aap, 456, 287 

\bibitem[Sood et al.(2007)]{sood07} Sood, R., Farrell, S., 
O'Neill, P., \& Dieters, S.\ 2007, Advances in Space Research, 40, 1528 

\bibitem[Terrell \& Wilson(2005)]{terrell05} Terrell, D., \& 
Wilson, R.~E.\ 2005, \apss, 296, 221 

\bibitem[Timmer \& Koenig(1995)]{tk95} Timmer, J., \& Koenig, M.\ 1995, \aap, 300, 707 

\bibitem[Titarchuk et al.(2007)]{tit07} Titarchuk, L., 
Shaposhnikov, N., \& Arefiev, V.\ 2007, \apj, 660, 556 

\bibitem[Thaller(1997)]{thaller97} Thaller, M.~L.\ 1997, \apj, 
487, 380 
\bibitem[Tomsick et al.(2006)]{tomsick06} Tomsick, J.~A., Chaty, 
S., Rodriguez, J., Foschini, L., Walter, R., \& Kaaret, P.\ 2006, \apj, 
647, 1309 

\bibitem[Vacca et al.(2003)]{vacca03} Vacca, W.~D., Cushing, 
M.~C., \& Rayner, J.~T.\ 2003, \pasp, 115, 389 

\bibitem[Vanbeveren et al.(1998)]{van98} Vanbeveren, D., de 
Donder, E., van Bever, J., van Rensbergen, W., \& de Loore, C.\ 1998, New 
Astronomy, 3, 443 

\bibitem[van der Hucht(2001)]{vanderhutch01} van der Hucht, K.~A.\ 
2001, New Astronomy Review, 45, 135 

\end{thebibliography}
\end{document}